\documentclass[12pt]{article}
\usepackage{amsmath}
\usepackage{amssymb}
\usepackage{graphicx}
\usepackage{epsfig}
\usepackage{cite}
\usepackage{url}
\usepackage[small]{caption}
\begin{document}
\baselineskip 0.6cm

\def\simgt{\mathrel{\lower2.5pt\vbox{\lineskip=0pt\baselineskip=0pt
           \hbox{$>$}\hbox{$\sim$}}}}
\def\simlt{\mathrel{\lower2.5pt\vbox{\lineskip=0pt\baselineskip=0pt
           \hbox{$<$}\hbox{$\sim$}}}}
\def\simprop{\mathrel{\lower3.0pt\vbox{\lineskip=1.0pt\baselineskip=0pt
             \hbox{$\propto$}\hbox{$\sim$}}}}
\def\bra#1{\langle #1 |}
\def\ket#1{| #1 \rangle}
\def\inner#1#2{\langle #1 | #2 \rangle}

\begin{titlepage}

\begin{flushright}
UCB-PTH-14/31\\
\end{flushright}

\vskip 1.5cm

\begin{center}
{\Large \bf Black Holes, Entropies, and Semiclassical Spacetime \\
in Quantum Gravity}

\vskip 0.7cm

{\large Yasunori Nomura and Sean J. Weinberg}

\vskip 0.4cm

{\it Berkeley Center for Theoretical Physics, Department of Physics,\\
 University of California, Berkeley, CA 94720, USA}

\vskip 0.1cm

{\it Theoretical Physics Group, Lawrence Berkeley National Laboratory,
 CA 94720, USA}

\vskip 0.8cm

\abstract{We present a coherent picture of the quantum mechanics of 
 black holes.  The picture does not require the introduction of any 
 drastically new physical effect beyond what is already known; it 
 arises mostly from synthesizing and (re)interpreting existing results 
 in appropriate manners.  We identify the Bekenstein-Hawking entropy 
 as the entropy associated with coarse-graining performed to obtain 
 semiclassical field theory from a fundamental microscopic theory of 
 quantum gravity.  This clarifies the issues around the unitary evolution, 
 the existence of the interior spacetime, and the thermodynamic nature 
 in black hole physics---any result in semiclassical field theory is 
 a statement about the maximally mixed ensemble of microscopic quantum 
 states consistent with the specified background, within the precision 
 allowed by quantum mechanics.  We present a detailed analysis of 
 information transfer in Hawking emission and black hole mining processes, 
 clarifying what aspects of the underlying dynamics are (not) visible 
 in semiclassical field theory.  We also discuss relations between 
 the black hole entropy and the entanglement entropy across the horizon. 
 We then extend our discussions to more general contexts in quantum 
 gravity.  The subjects include extensions to de~Sitter and Minkowski 
 spaces and implications for complementarity and cosmology, especially 
 the eternally inflating multiverse.}

\end{center}
\end{titlepage}

\section{Introduction}
\label{sec:intro}

One of the major cornerstones in the pursuit of a quantum theory of 
gravity was the discovery of the finite nonzero entropy of a black 
hole~\cite{Bekenstein:1973ur} and the associated thermal radiation from 
it~\cite{Hawking:1974sw}.  The precise interpretation of this result, 
however, has not been entirely clear.  What does this entropy represent? 
Where does it reside?  The fact that a black hole emits radiation at the 
quantum level allowed us to contemplate the possibility that a complete 
description of the system is obtained by referring only to the spacetime 
region outside the horizon; in particular, the evolution of a black 
hole can be unitary when described from a distance~\cite{'tHooft:1990fr}. 
The existence of the interior spacetime then becomes manifest 
only after changing the description to an infalling one, which 
is supposed to provide a description complementary to the distant 
one~\cite{Susskind:1993if,Hayden:2007cs}.  This last picture, however, 
has recently been challenged~\cite{Almheiri:2012rt}:\ if the emission 
process is indeed unitary, then an infalling observer is claimed to 
encounter something drastic at the horizon, so that there is no such 
thing as the interior spacetime, at least for an old black hole.

The main purpose of this paper is to present a coherent picture of the 
quantum mechanics of black holes and to clarify the issues around their 
unitary evolution and the existence of the interior spacetime, although 
we also extend our discussions to more general contexts in quantum gravity. 
We adopt the hypothesis that, as postulated in Ref.~\cite{'tHooft:1990fr}, 
there is a unitary description of a black hole which involves only the 
region outside the horizon.  We discuss how this hypothesis is consistent 
with the existence of the interior spacetime, in the sense of complementarity 
in Ref.~\cite{Susskind:1993if}.  Our picture does not involve any 
drastically new physical effect beyond what is already known or postulated, 
and yet it requires a certain revision of the applicability of semiclassical 
theory in discussing physics around the black hole.  We feel that this 
provides a significant advancement toward understanding the structure 
of quantum gravity and its relation to the emergent semiclassical 
picture of spacetime.

Our discussion begins with studying the origin of the Bekenstein-Hawking 
entropy.  In general, the concept of entropy is associated with 
coarse-graining.  What is the coarse-graining responsible for the 
Bekenstein-Hawking entropy?  Since the Bekenstein-Hawking entropy 
arises in semiclassical field theory without doing any coarse-graining 
in the theory, it must be associated with the coarse-graining performed 
{\it to obtain the theory} from a fundamental theory of quantum gravity. 
This picture is indeed consonant with the fact that in quantum mechanics, 
having a well-defined geometry of spacetime, e.g.\ a black hole formed 
in a well-defined spacetime location, requires taking a superposition 
of an enormous number of energy-momentum eigenstates, so we expect 
that there are many different ways to arrive at the same background 
for a semiclassical field theory within the precision allowed by 
quantum mechanics.  In particular, this implies that detailed microscopic 
physics occurring within the fine-grained degrees of freedom, including 
the flow of microscopic information in the Hawking emission process, 
cannot be seen in semiclassical field theory.  This is the origin of 
the apparent violation of unitarity~\cite{Hawking:1976ra} in the Hawking 
evaporation process.

The fact that the microstates of a black hole cannot be resolved in 
semiclassical field theory implies that for the purpose of describing 
physics of excitations in semiclassical field theory---including 
an ordinary observer falling into the black hole---we may ignore 
the dynamics within the fine-grained microscopic degrees of freedom, 
which is responsible for the fundamental unitarity of the Hawking process. 
Specifically, we may view that a semiclassical field theory vacuum 
has a hidden ``vacuum index'' $k$ labeling the microstates, and can 
construct field theory operators that act in the same way on each of 
these exponentially many vacuum states:\ $k = 1,\cdots,e^S$, where $S$ 
represents the Bekenstein-Hawking entropy.  This allows us to construct 
operators describing the interior spacetime following the standard 
method in semiclassical field theory~\cite{Unruh:1976db}.  Note that 
this does not contradict the hypothesis that the evaporation of a black 
hole {\it when described in the fundamental theory} is a unitary process.

Since emissions of later Hawking quanta occur within the causal future 
of the region in which earlier Hawking emissions have occurred, there 
is no reason to expect that the information recovery from a black 
hole must violate locality, or causality, in spacetime as described 
in a distant reference frame.  We perform a detailed analysis of the 
Hawking emission process, as well as black hole mining, and argue that 
the modes of a black hole relevant for the information transfer are 
not visible to semiclassical field theory---they are ``too soft'' to 
be resolved.  In semiclassical field theory, Hawking emission occurs 
because of an intrinsic quantum mechanical ambiguity of defining 
particles in curved spacetime, and the information about the microstate 
of a black hole it carries away is viewed as being stored nonlocally 
in states representing the field theory vacuum, which extends into the 
whole zone region.  Note that this does not contradict the locality 
{\it of the dynamics}; in quantum mechanics, the information about 
a state can in general be delocalized even if the dynamics is local. 
This also implies that the purifiers of the emitted Hawking quanta 
are what semiclassical gravity describes as vacuum states; in 
particular, they are not outgoing field theory modes as envisioned 
in Refs.~\cite{Almheiri:2012rt,Almheiri:2013hfa}.

The picture described in this paper has been developed over the past 
years in Refs.~\cite{Nomura:2012ex,Nomura:2013lia}, although its 
final form presented here does not agree in all respects with those 
given in these earlier works. Following the inspirational work in 
Ref.~\cite{Almheiri:2012rt}, there appeared numerous attempts to 
reconcile the unitary evolution of a black hole with the existence 
of the interior, some of which introduce dramatic new physics; 
see, e.g., Refs.~\cite{Giddings:2012bm,Papadodimas:2012aq,%
Verlinde:2012cy,Maldacena:2013xja,Bousso:2012as}.  To give 
examples, Giddings~\cite{Giddings:2012bm} studied the possibility 
that the dynamics in the exterior region may be nonlocal. 
Papadodimas and Raju~\cite{Papadodimas:2012aq} and Verlinde 
and Verlinde~\cite{Verlinde:2012cy} considered ``state-dependent'' 
maps to describe the interior region.  Maldacena and 
Susskind~\cite{Maldacena:2013xja} considered the possibility 
that a part of the degrees of freedom describing the interior 
region comes from Hawking radiation emitted earlier.  Our picture 
does not require the introduction of such new physics.

The organization of this paper is as follows.  In Section~\ref{sec:entropy}, 
we discuss how to interpret the Bekenstein-Hawking entropy, in particular 
the way it may manifest when the black hole interacts with other systems. 
In Section~\ref{sec:semiclass}, we discuss how the microscopic structure 
of quantum gravity is related to the semiclassical view of the world. 
We argue that semiclassical field theory corresponds to the description 
arising after coarse-graining the degrees of freedom associated with 
the Bekenstein-Hawking entropy, specifically after taking the maximally 
mixed ensemble of microstates consistent with the given background 
within some precision.  This explains various confusing aspects of 
black hole physics regarding fundamental unitarity and the thermodynamic 
nature of semiclassical gravity.  We also discuss relations between 
the black hole entropy and the entanglement entropy across the horizon. 
In Section~\ref{sec:micro}, we present a detailed analysis of the 
information transfer in the Hawking emission and black hole mining 
processes, especially focusing on what aspects of the underlying 
dynamics can be captured in semiclassical field theory. Finally, 
in Section~\ref{sec:discuss}, we extend our discussions to more general 
contexts in quantum gravity, including discussions of de~Sitter and 
Minkowski spaces and implications for complementarity and cosmology, 
especially the eternally inflating quantum multiverse.

Throughout the paper, we adopt the Schr\"{o}dinger picture for quantum 
evolution, and adopt natural units in which $\hbar = c = 1$, unless 
otherwise stated.  $l_{\rm P} \simeq 1.62 \times 10^{-35}~{\rm m}$ 
denotes the Planck length.  In our discussion of black hole physics, 
we focus on a black hole that is well approximated by a Schwarzschild 
black hole in 4-dimensional asymptotically flat spacetime.  We do not 
expect difficulty in extending it to other more general cases.

\section{How to Interpret the Black Hole Entropy?}
\label{sec:entropy}

In this section, we discuss an interpretation of the black hole entropy, 
including how it manifests when the black hole interacts with other 
systems.  Understanding this issue correctly is important in the development 
of our picture in later sections.

\subsection{Bekenstein-Hawking entropy from spatial localization}
\label{subsec:density}

We first remind the readers that the Bekenstein-Hawking entropy is an 
{\it entropy density}.  It is a finite function of the black hole mass 
$M$ because we restrict our considerations to a black hole in a specific 
spacetime region.

Let us begin our discussion by considering a simple quantum system 
consisting of massless particles that do not have a conserved charge. 
The quantum states representing a single particle can be labeled by three 
continuous numbers corresponding to the momentum of the particle ${\bf k} 
= (k_x,k_y,k_z)$, so that the states with $N$ (weakly interacting) particles 
are labeled by $3N$ continuous numbers.  Consider now a set of quantum 
states $\ket{\psi_i}$ in which all the particles in the system are 
confined in a spatial region with volume $V$ at some time $t = t_0$. 
These states are obtained (only) by taking appropriate superpositions 
of states in which particles have well-defined momenta.  How many such 
states exist?  Assuming that the system has energy between $E$ and 
$E + \delta E$, quantum mechanics tells us that the number of independent 
$\ket{\psi_i}$ states is finite and of order ${\cal N} \approx 
e^{O(\rho^{3/4} V)} \delta E/E$, where $\rho \equiv E/V$.  Namely, 
the entropy density is $S = \ln ({\cal N} E/\delta E) \approx 
O(\rho^{3/4} V)$, which does not depend on how $\delta E$ is chosen 
as long as $\delta E/E \gg e^{-S}$.  Note that physics described 
here is intrinsically quantum mechanical in the sense that classical 
mechanics would allow the specified spatial region to support a 
continuously infinite number of states with energy between $E$ and 
$E + \delta E$.  This can be seen from the fact that $S$ can be written 
as $O(\rho^{3/4} V/\hbar^{3/4} c^{3/4})$ when $\hbar$ and $c$ are 
restored, so that $S \rightarrow \infty$ for $\hbar \rightarrow 0$.

The finiteness of the Bekenstein-Hawking entropy suggests that the 
situation for a black hole is similar.  Let us consider forming 
a black hole of mass $M$ by collapsing matter.  In order to have such 
a process in which the black hole is formed at a well-defined spacetime 
location, the initial matter state must involve a superposition of 
energy-momentum eigenstates.  Suppose we want to identify the spacetime 
location of the black hole with precision comparable to the quantum 
stretching of the horizon $\varDelta r \approx O(1/M)$, i.e.\ $\varDelta d 
\approx O(l_{\rm P})$, and the timescale of Hawking emission $\varDelta t 
\approx O(M l_{\rm P}^2)$, where $r$ and $t$ are the Schwarzschild radial 
and time coordinates, respectively, and $d$ is the proper length.  In this 
case the superposition must involve momenta with spread $\varDelta p \simgt 
O(1/M l_{\rm P}^2)$ and energy with $\varDelta E \simgt O(1/M l_{\rm P}^2)$, 
where $\varDelta p$ and $\varDelta E$ are both measured in the asymptotic 
region, and these uncertainties must exist shortly before the formation 
of the black hole in the branch we are interested in.  For an older 
black hole, the stochastic nature of Hawking radiation introduces 
macroscopic uncertainties for the black hole mass, location, and 
spin~\cite{Nomura:2012cx,Page:1979tc}.  This effect, however, can 
be easily separated by focusing on appropriate branches in the full 
quantum state.  A consideration similar to the one above then applies 
in the limit that the induced spin of the black hole is neglected, 
which is a reasonable approximation for a large black hole.

How many black hole states do there exist in which the black hole is 
at a specific location within an uncertainty of $\varDelta r \approx O(1/M)$ 
and $\varDelta t \approx O(M l_{\rm P}^2)$ measured in the asymptotic region? 
Because of the required uncertainty in energy-momentum, we must consider 
the number of black hole states of mass (which we may identify with 
the energy) in the range between $M$ and $M + \delta M$ where $\delta M 
\simgt O(1/M l_{\rm P}^2)$.  As in the case before, quantum mechanics 
makes the number of independent such localized states finite.  Indeed, 
the logarithm of this number---the entropy density---is given by the 
Bekenstein-Hawking formula
\begin{equation}
  S = \frac{{\cal A}}{4 l_{\rm P}^2},
\label{eq:S_BH}
\end{equation}
at the leading order in expansion in inverse powers of ${\cal A}/l_{\rm P}^2$, 
where ${\cal A} = 16\pi M^2 l_{\rm P}^4$ is the area of the horizon.  Here 
we identify the expression of Eq.~(\ref{eq:S_BH}) to represent the density 
of black hole {\it vacuum} states, i.e.\ states representing a black hole 
that does not have an excitation in the interior or the exterior region. 
This is reasonable because the Bekenstein-Hawking considerations apply to 
a black hole after the horizon is classically stabilized.  Possible higher 
order corrections to Eq.~(\ref{eq:S_BH}) do not affect our argument, so 
we will ignore them.  The states in which there are excitations in the 
interior or the near exterior region will be discussed below.

As can be easily seen, the expression in Eq.~(\ref{eq:S_BH}) is insensitive 
to the precise choice of $\delta M$ as long as $\delta M/M \gg e^{-S}$. 
For definiteness, we will mostly take the smallest possible value of 
$\delta M$:
\begin{equation}
  \delta M \approx O\biggl(\frac{1}{M l_{\rm P}^2}\biggr).
\label{eq:delta-M}
\end{equation}
The corresponding uncertainty in timescale is then $\varDelta t \approx 
1/\delta M \approx O(M l_{\rm P}^2)$.  Only an $O(1)$ number of Hawking 
quanta with energy $E \approx O(1/M l_{\rm P}^2)$ are emitted within 
$\varDelta t$, so a black hole does not change its mass more than 
$O(\delta M)$ in this timescale.

The discussion above demonstrates that the finite entropy of a black hole 
can be understood in a similar manner to that of other quantum systems 
confined in a finite spatial region, although there are important 
differences including the fact that the black hole entropy scales 
as the area unlike those of usual (much) lower density materials which 
scale as the volume $V$ for fixed intensive quantities, e.g., $\rho$.%
\footnote{Following standard convention, here and below we use the term 
 entropy to also mean entropy density.}
(This leads to the conjecture for the universal entropy bound associated with 
the area of a codimension-2 surface~\cite{'tHooft:1993gx,Susskind:1994vu}; 
a precise formulation of it must involve null hypersurfaces rather 
than spatial regions~\cite{Bousso:1999xy}.)  In particular, it is not 
appropriate to consider that quantum mechanics introduces an exponentially 
large number of degeneracies for the microstates that do not exist in 
the corresponding classical black hole.  In classical general relativity, 
a set of Schwarzschild black holes located at some place at rest are 
parameterized by a continuous mass parameter $M$; i.e., there are a 
continuously infinite number of black hole states in the energy interval 
between $M$ and $M + \delta M$ for any $M$ and small $\delta M$.  Quantum 
mechanics {\it reduces} this to a finite number $\approx e^S \delta M/M$ 
with $S$ given by Eq.~(\ref{eq:S_BH}).%
\footnote{Of course, quantum mechanics allows for a superposition of 
 these finite number of independent states, so the number of possible 
 (not necessarily independent) states is continuously infinite.  The 
 statement here applies to the number of independent states, regarding 
 classical black holes with different $M$ as independent states.}
This can also be seen from the fact that $S$ is written as ${\cal A} c^3/4 
l_{\rm P}^2 \hbar$ when $\hbar$ and $c$ are restored, which becomes 
infinite for $\hbar \rightarrow 0$.

We now discuss black hole states having excitations in the interior 
or near exterior region.  (In a distant reference frame, an excitation 
in the interior region is described as an excitation of the horizon.) 
Consider the set $C_I$ of the states which have specified excitations 
$I$ in a background of a black hole of mass $M$ (more precisely between 
$M$ and $M + \delta M$).  As suggested, e.g., by a representative estimate 
in Ref.~\cite{'tHooft:1993gx}, the entropy associated with all these sets 
is expected to be written roughly as
\begin{equation}
  S = \frac{{\cal A}}{4 l_{\rm P}^2} + \sum_I S_I;
\qquad
  \sum_I |S_I| \approx 
    O\biggl( \frac{{\cal A}^q}{l_{\rm P}^{2q}};\, q < 1 \biggr),
\label{eq:S_excit}
\end{equation}
where ${\cal A} = 16\pi M^2 l_{\rm P}^4$, and $S_I$ is the entropy associated 
with the class of excitations $I$.  The total entropy $S$, therefore, 
is simply ${\cal A}/4 l_{\rm P}^2$ up to fractional corrections in inverse 
powers of ${\cal A}/l_{\rm P}^2$.  Namely, the existence of excitations 
provides only small perturbation in terms of the entropy counting.  (This 
point was particularly emphasized in Refs.~\cite{Nomura,Nomura:2013lia}.) 
Note that being defined as fluctuations with respect to a fixed background 
spacetime, the energy $E_I$ and entropy $S_I$ of excitations can be either 
positive or negative.

The approximate nature of Eq.~(\ref{eq:S_excit}) needs to be emphasized. 
First, dividing a physical configuration into excitations and background 
is artificial.  In the expression in Eq.~(\ref{eq:S_excit}), we have simply 
added the entropies of the excitations to that of the black hole.  This is 
appropriate since we can view semiclassical physics occurring in a fixed 
black hole background to correspond to excitations on {\it each} of the 
black hole vacuum states (see Section~\ref{subsec:interior}).  On the other 
hand, it is clear that this treatment involves an approximation---because 
the existence of excitations must affect geometry, the ``on-shell'' 
Hilbert space (the space of states after the equations of motion are 
imposed) must take the form more complicated than the one implied by 
Eq.~(\ref{eq:S_excit}).  Nevertheless, Eq.~(\ref{eq:S_excit}) represents 
the way semiclassical field theory treats physical systems, which, with 
some care, may capture certain aspects of quantum gravitational physics.

\subsection{Where does the black hole entropy reside?}
\label{subsec:where}

Consider the set of all the independent black hole vacuum states 
$\ket{\Psi_k(M)}$ of mass between $M$ and $M + \delta M$, with the black 
hole localized in a specified spatial region with precision $\varDelta r 
\approx O(1/M)$.  As discussed in the previous subsection, the index 
$k$ runs over
\begin{equation}
  k = 1, \cdots, e^S = e^{4\pi M^2 l_{\rm P}^2} \equiv n(M),
\label{eq:k}
\end{equation}
representing the Bekenstein-Hawking entropy.  Where does this entropy reside?

Let us assume that physics is local in the region outside the stretched 
horizon, in particular we can use appropriately quantized Einstein gravity 
to calculate physical quantities within its regime of validity.  It is 
important to remind ourselves that locality is a property of dynamics, in 
particular the Hamiltonian, and not that of states.  In fact, in quantum 
mechanics the information about a state is quite generally delocalized in 
space.  Consider, for instance, states $\ket{1}$ and $\ket{2}$ representing, 
respectively, a particle of species $1$ and $2$ in a momentum eigenstate, 
located in a finite box with periodic boundary conditions.  Suppose the 
system is in one of these two states.  Where does the information about 
the state exist?  The answer is:\ everywhere.  We may find out if the 
system is in $\ket{1}$ or $\ket{2}$ anywhere in the box, as long as we 
are equipped with an appropriate detector and have enough time for the 
measurement.%
\footnote{This does not mean that there is no physical information flow 
 associated with the measurement process.  It is simply that to discuss 
 the information flow, we need to consider the process preparing the 
 state.  For example, if the particle is created at some point (e.g.\ 
 in the form of a wavepacket, which later broadens), the information 
 must be regarded as being transferred from the creation point to the 
 detector location in the branch in which the detector has responded. 
 If the dynamics is local, this information transfer does not violate 
 causality in spacetime.}

Quite analogously, we assume that a part of the information about the 
index $k$ of a black hole vacuum state is delocalized over a large spatial 
region.  In particular, we consider that this occurs mostly in the region 
$r \simlt 3M l_{\rm P}^2$, which, in a distant view, consists of the 
stretched horizon at $r = r_{\rm s} = 2Ml_{\rm P}^2 + O(1/M)$ and 
near exterior ``zone'' $r_{\rm s} < r \simlt 3M l_{\rm P}^2$, the region 
inside the effective gravitational potential barrier separating the near 
and far exterior regions.  To isolate the part of the state that carries 
a meaningful amount of information, let us separate the Hilbert space 
into two factors:\ one representing states in the region $r \leq R$, 
${\cal H}_<$, and the other $r > R$, ${\cal H}_>$.  Here, the separation 
radius $R$ may be conveniently chosen to be somewhere in the outer side of 
the potential barrier, e.g.\ in the range $(3~\mbox{--}~5) M l_{\rm P}^2$. 
The black hole vacuum state $\ket{\Psi_k(M)}$ can then be written 
approximately as
\begin{equation}
  \ket{\Psi_k(M)} = \sum_a c_a \ket{\psi_{ka}(M)} \ket{\phi_a(M)},
\label{eq:separate}
\end{equation}
where $\ket{\psi_{ka}(M)}$ and $\ket{\phi_a(M)}$ are elements of 
${\cal H}_<$ and ${\cal H}_>$, respectively.  The statement that the 
information is spread essentially only in the region $r_{\rm s} \leq 
r \simlt 3M l_{\rm P}^2$ corresponds to the absence of the index $k$ 
for $c_a$ and $\ket{\phi_a(M)}$.

We note that by the information delocalization above, we do not mean 
that the information is spread {\it uniformly} over the region $r_{\rm s} 
\leq r \simlt 3M l_{\rm P}^2$.  In fact, we expect that most of the 
information is localized in the region close to the horizon.  Specifically, 
when we split the Hilbert space at $r = R' < R$, we have the expression
\begin{equation}
  \ket{\Psi_k(M)} = \sum_a c'_a \ket{\psi'_{ia}(M)} \ket{\phi'_{ja}(M)},
\label{eq:separate'}
\end{equation}
where $k = \{i,j\}$, and $i$ and $j$ take values $i = 1,\cdots,I$ and 
$j=1,\cdots,J$ with $IJ = n(M)$.  We expect that $s(R') \equiv \ln J$, 
which is a function of $R'$, is given by the thermal entropy contained 
in the region $r > R'$, calculated in semiclassical field theory using 
the blueshifted local Hawking temperature.  In particular,
\begin{equation}
  s(R') \ll \ln n(M),
\label{eq:weak}
\end{equation}
unless $R'-2M l_{\rm P}^2 \simlt O(1/M)$.  Note that taking $J=1$, which 
corresponds to $R' \rightarrow R$, Eq.~(\ref{eq:separate'}) is reduced 
to Eq.~(\ref{eq:separate}).  An important point here is that {\it some} 
information about $k$ can be extracted by a physical process occurring 
as far as $r \simeq 3M l_{\rm P}^2$ in timescale of order $1/\delta M$ 
(more precisely, without directly interacting with the stretched horizon 
at the field theory level; see Section~\ref{sec:micro} for further 
discussions).

From now on, we suppress the trivial entanglement in Eq.~(\ref{eq:separate}) 
that does not depend on $k$ for the simplicity of the notation.  This 
allows us to write the black hole vacuum state $\ket{\Psi_k(M)}$ (very 
roughly) as
\begin{equation}
  \ket{\Psi_k(M)} \approx \ket{\psi_k(M)} \ket{0_{\rm ext}},
\label{eq:separate-2}
\end{equation}
where $\ket{0_{\rm ext}}$ is the vacuum state in the region $r > R$, 
which does not depend on $k$.  While $\ket{0_{\rm ext}}$ has some 
dependence on $M$, for our purposes here we will ignore it.  We have 
also assumed the absence of Hawking quanta emitted at earlier times. 
If there are excitations in the interior region (or of the stretched 
horizon in a distant view) or in the zone, then $\ket{\psi_k(M)}$ in 
Eq.~(\ref{eq:separate-2}) must be replaced with the corresponding excited 
states $\ket{\tilde{\psi}_{kn}(M)}$, where $n$ labels the excitations. 
The dimension of the relevant Hilbert spaces satisfies
\begin{equation}
  \ln {\rm dim}\,{\cal H}_{\tilde{\psi}_k} 
  \approx O\Bigl( M^{2q} l_{\rm P}^{2q};\, q < 1 \Bigr),
\label{eq:dim-H_tilde-psi}
\end{equation}
where ${\cal H}_{\tilde{\psi}_k}$ is the Hilbert space spanned by 
$\ket{\psi_k(M)}$ and $\ket{\tilde{\psi}_{kn}(M)}$ for a fixed $k$, 
and all the ${\cal H}_{\tilde{\psi}_k}$'s with different $k$ (as well 
as the structure of field theory operators defined for each of them) are 
isomorphic with each other; see Section~\ref{subsec:interior}.  If the 
far exterior region is not in the vacuum, the state $\ket{0_{\rm ext}}$ 
in Eq.~(\ref{eq:separate-2}) must be replaced accordingly.

How can we probe the microscopic information associated with a black 
hole vacuum state?  Consider a physical detector located somewhere in 
the region $r_{\rm s} < r \simlt 3M l_{\rm P}^2$ whose ground state 
$\ket{d_0}$ represents the ``ready'' state while excited states 
$\ket{d_i}$ ($i=1,2,\dots$) are the pointer states.  Suppose the 
proper energies needed to excite $\ket{d_0}$ to $\ket{d_i}$ are given 
by $E_{{\rm d},i}$.  The state representing the region $r \leq R$ 
then evolves as
\begin{equation}
  \ket{\psi_k(M)} \ket{d_0} \rightarrow \sum_i \sum_{k_i = 1}^{n(M_i)} 
    \alpha^k_{k_i i} \ket{\psi_{k_i}(M_i)} \ket{d_i};
\qquad
  M_i = M - E_{{\rm d},i} \sqrt{1-\frac{2M l_{\rm P}^2}{r_{\rm d}}},
\label{eq:probe}
\end{equation}
where the function $n(M)$ is defined in Eq.~(\ref{eq:k}).  The coefficients 
$\alpha^k_{k_i i}$ in general depend on the location of the detector, and 
$M_i$ for different $i$ may belong to the same mass within the precision 
$\delta M$, i.e.\ $M_i = M_{i'}$ for $i \neq i'$.  Here, we have separated 
the detector state from the rest of the system, although in a complete 
treatment the detector itself may be better viewed as an excitation 
over $\ket{\psi_k(M)}$.

The process in Eq.~(\ref{eq:probe}) implies that (a part of) the information 
encoded in the index $k$ can be probed by the detector.  In fact, the 
detailed microscopic process leading to Eq.~(\ref{eq:probe}) is somewhat 
more subtle.  In particular, in Eq.~(\ref{eq:probe}) we have assumed 
that the black hole states appearing in the right-hand side are vacuum 
states, but this is the case only after multiple elementary processes 
have occurred.  We will discuss these processes in Section~\ref{sec:micro}.

Emission of Hawking quanta to the asymptotic region also carries 
information away from the black hole.  Since the effective gravitational 
potential is damped in the region $r \simgt R$, Hawking quanta emitted 
from the region $r \sim R$ propagate essentially freely to the asymptotic 
region (except that they receive a small residual gravitational redshift 
of a factor of about $1.5$).  The elementary emission process may thus be 
written as
\begin{equation}
  \ket{\psi_k(M)} \ket{0_{\rm ext}} \rightarrow 
    \sum_i \sum_{k_i = 1}^{n(M-E_i)} 
    \beta^k_{k_i i} \ket{\psi_{k_i}(M-E_i)} \ket{i_{\rm ext}},
\label{eq:emission}
\end{equation}
where $\ket{i_{\rm ext}}$ represents a state with energy $E_i$ in which 
outgoing radiation modes are excited.  Again, we have put black hole 
vacuum states after the evolution in the right-hand side; the eligibility 
of this will be discussed in Section~\ref{sec:micro}.  Since the final 
state depends on $k$, this evolution can be unitary, which we assume 
to be the case.

We stress that the emission process in Eq.~(\ref{eq:emission}) can be 
viewed as occurring locally in the potential barrier region because of 
the information delocalization discussed above.  To elucidate this point, 
we consider the tortoise coordinate
\begin{equation}
  r^* = r + 2M l_{\rm P}^2\, \ln \frac{r-2M l_{\rm P}^2}{2M l_{\rm P}^2},
\label{eq:tortoise}
\end{equation}
in which the region outside the Schwarzschild horizon $r \in (2M l_{\rm P}^2, 
\infty)$ is mapped into $r^* \in (-\infty,\infty)$.  This coordinate 
is useful in that the kinetic term of an appropriately redefined field 
takes the canonical form, so that its propagation can be analyzed as 
in flat space.  In this coordinate, the stretched horizon, located at 
$r = 2M l_{\rm P}^2 + O(l_{\rm s}^2/M l_{\rm P}^2)$, is at
\begin{equation}
  r^*_{\rm s} \simeq -4M l_{\rm P}^2 \ln\frac{M l_{\rm P}^2}{l_{\rm s}},
\label{eq:rs_stretched}
\end{equation}
where $l_{\rm s}$ is the string (cutoff) scale, which we take to be within 
a couple of orders of magnitude of $l_{\rm P}$.  This implies that there 
is a large distance between the stretched horizon and the potential barrier 
region when measured in $r^*$:\ $\varDelta r^* \approx 4M l_{\rm P}^2 
\ln(M l_{\rm P}) \gg O(M l_{\rm P}^2)$ for $\ln(M l_{\rm P}) \gg 1$. 
On the other hand, a localized Hawking quantum is represented by a 
wavepacket with width of $O(M l_{\rm P}^2)$ in $r^*$, since it has 
an energy of order $T_{\rm H} = 1/8\pi M l_{\rm P}^2$ defined in the 
asymptotic region.

The point is that, given the black hole state $\ket{\psi_k(M)}$, the 
process in Eq.~(\ref{eq:emission}) occurs in the region $|r^*| \approx 
O(M l_{\rm P}^2)$ without involving deep interior of the zone $-r^* 
\gg M l_{\rm P}^2$.  In this region, the information stored (nonlocally) 
in the vacuum state is converted into that of a particle state outside 
the zone, where the concept of particles with frequencies $\omega 
\simlt 1/M l_{\rm P}^2$ is well defined.  A corollary of this statement 
is that if we evolve the system backward in time, an originally outgoing 
Hawking quantum does not become a highly blueshifted ingoing quantum 
in a region deep in the zone.  Instead, it becomes ``vacuum degrees 
of freedom,'' $k$, due to interactions with spacetime caused by the 
curvature.  Since the concept of particles with $\omega \simlt 1/M 
l_{\rm P}^2$ is not well defined in the zone region, we need not view 
this as new physics contradicting conventional low energy Einstein gravity. 
It is simply that semiclassical field theory cannot be used to analyze 
these quanta within the zone, where the scale of curvature length is 
of order $M l_{\rm P}^2$.

\section{What Is the Semiclassical Approximation?}
\label{sec:semiclass}

In this section, we discuss relations between the picture presented in 
the previous section and the standard semiclassical treatment of a black 
hole.  This will clarify the meaning of the Bekenstein-Hawking entropy 
further, and shed important light on the nature of the semiclassical 
approximation in quantum gravity.

It is often stated that in the semiclassical approximation, in 
particular quantum field theory on a fixed spacetime background, particle 
excitations are treated quantum mechanically while the background spacetime 
classically.  The interpretation of this statement, however, needs care. 
As was seen in the previous section, if a black hole is treated ``fully 
classically,'' its entropy is infinite, and so it has zero temperature 
(and hence does not radiate).  On the other hand, Hawking's calculation 
finds a nonzero temperature in semiclassical gravity~\cite{Hawking:1974sw}. 
This implies that, as in any statistical mechanical understanding 
of entropy (which requires, e.g., quantization of a phase space), 
semiclassical gravity is capturing certain, though not all, quantum 
aspects of the relevant physical system, in this case the background 
spacetime.

So, what is the semiclassical approximation really?  We assert that 
any result in the semiclassical approximation is a statement about 
the (maximally mixed) {\it ensemble} of microscopic quantum states 
consistent with the specified spacetime, within the precision allowed 
by quantum mechanics.  In the case of a black hole at a fixed location, 
the semiclassical approximation deals with the mixed state
\begin{equation}
  \rho(M) = \frac{1}{n(M)} 
    \sum_{k=1}^{n(M)} \ket{\Psi_k(M)} \bra{\Psi_k(M)}.
\label{eq:rho_M}
\end{equation}
Because the relevant density matrix $\rho$ in general takes the form 
of a maximally mixed state, results in semiclassical gravity do not 
depend on the basis of the microscopic pure states chosen, in the black 
hole case of the $\ket{\Psi_k(M)}$'s.  Below we will elucidate the above 
assertion by considering the thermal nature (in Section~\ref{subsec:thermal}) 
and the interior spacetime (in Section~\ref{subsec:interior}) of a 
semiclassical black hole.

By construction, the semiclassical approximation cannot capture unitarity 
of detailed microscopic processes associated with spacetime because 
it involves coarse-graining in the sense of Eq.~(\ref{eq:rho_M}). 
This is why Hawking's calculation found apparent violation of unitarity 
in the black hole evaporation process~\cite{Hawking:1976ra}.  In this 
section, we limit our discussion to the level in which the details of 
these microscopic processes are ignored.  The underlying microscopic 
dynamics of quantum gravity and its implications for the semiclassical 
approximation will be discussed further in Section~\ref{sec:micro}.

\subsection{Thermal nature}
\label{subsec:thermal}

According to the semiclassical calculation, a black hole of mass $M$ 
emits black-body radiation, corrected by gray-body factors, of temperature 
$T_{\rm H} = 1/8\pi M l_{\rm P}^2$.  In particular, if a detector is 
located at $r = r_{\rm d}$, then it will see blueshifted Hawking radiation 
with temperature
\begin{equation}
  T(r_{\rm d}) = \frac{1}{8\pi M l_{\rm P}^2 
   \sqrt{1-\frac{2Ml_{\rm P}^2}{r_{\rm d}}}}.
\label{eq:T_rad}
\end{equation}
What does this mean at the microscopic level?

Suppose we put the model detector considered in Section~\ref{subsec:where} 
somewhere in the zone, $r_{\rm s} < r_{\rm d} \simlt 3M l_{\rm P}^2$. 
Specifically, the detector has the ready state $\ket{d_0}$ and pointer 
states $\ket{d_i}$ ($i = 1,2,\cdots$) with the proper energies needed 
to excite $\ket{d_0}$ to $\ket{d_i}$ given by $E_{{\rm d},i}$.  Applying 
Eq.~(\ref{eq:probe}) (with the understanding that we are focusing on 
the branch in which the detector has responded, and that the right-hand 
side is normalized), the interaction between the detector and the mixed 
state in Eq.~(\ref{eq:rho_M}) is given by
\begin{equation}
  \rho(M) \otimes \ket{d_0} \bra{d_0} 
  \rightarrow 
    \frac{1}{n(M)} \sum_{k=1}^{n(M)}\; \sum_{i,i'} 
    \sum_{k_i = 1}^{n(M_i)} \sum_{k'_{i'} = 1}^{n(M_{i'})} 
    \alpha^k_{k_i i} \alpha^{k*}_{k'_{i'} i'} 
    \ket{\psi_{k_i}(M_i)} \ket{d_i} 
    \bra{\psi_{k'_{i'}}(M_{i'})} \bra{d_{i'}}.
\label{eq:int-mixed}
\end{equation}
This leads to the density matrix describing the detector state after 
the interaction
\begin{equation}
  \rho_{\rm d} = \sum_{i,i'} \gamma_{ii'}\, \ket{d_i} \bra{d_{i'}};
\qquad
  \gamma_{ii'} = \frac{1}{n(M)} \sum_{k=1}^{n(M)} \sum_{k_i=1}^{n(M_i)} 
    \alpha^k_{k_i i} \alpha^{k*}_{k_i i} \delta_{ii'},
\label{eq:rho_d}
\end{equation}
where we have assumed $M_i \neq M_{i'}$ for $i \neq i'$ for simplicity. 
The result of the semiclassical calculation implies that this density 
matrix takes the form as if the detector is immersed in the thermal 
bath of temperature in Eq.~(\ref{eq:T_rad}).  If the sensitivity of 
the detector does not depend on the excitation level $i$,
\begin{equation}
  \gamma_{ii'} \approx \frac{1}{Z} 
    e^{-\frac{E_{{\rm d},i}}{T(r_{\rm d})}}\, \delta_{ii'},
\label{eq:therm}
\end{equation}
where $Z = \sum_i e^{-E_{{\rm d},i}/T(r_{\rm d})}$.  Note that while 
the fundamental process in Eq.~(\ref{eq:probe}) leads to the correlation 
between the detector and the microstate of the black hole $k$, the 
density matrix $\rho_d$ in Eq.~(\ref{eq:rho_d}) does not reflect it, 
because of the maximally mixed nature of the state $\rho(M)$.

The situation in Hawking emission, in this respect, is similar.  As 
discussed around Eq.~(\ref{eq:emission}), the fundamental emission 
process, occurring in a time interval of order $1/\delta M \approx 
M l_{\rm P}^2$, can be written as
\begin{equation}
  \ket{\psi_k(M)} \ket{\phi_a} \rightarrow 
    \sum_i \sum_{k_i = 1}^{n(M-E_i)} 
    \beta^k_{k_i i} \ket{\psi_{k_i}(M-E_i)} \ket{\phi_{a+i}},
\label{eq:emission-gen}
\end{equation}
where $\ket{\phi_a}$ is a general far exterior state representing the 
region $r > R$, and $\ket{\phi_{a+i}}$ is the state in which newly emitted 
Hawking quanta, labeled by $i$ and having total energy $E_i$, are added 
to the appropriately time evolved $\ket{\phi_a}$.  Applying this to the 
``semiclassical state'' in Eq.~(\ref{eq:rho_M}), i.e.
\begin{equation}
  \rho(M) = \frac{1}{n(M)} \sum_{k=1}^{n(M)} 
    \ket{\psi_k(M)} \ket{\phi_a} \bra{\psi_k(M)} \bra{\phi_a},
\label{eq:rho_M-2}
\end{equation}
its evolution is given by
\begin{equation}
  \rho(M) \rightarrow 
    \frac{1}{n(M)} \sum_{k=1}^{n(M)}\; \sum_{i,i'} 
    \sum_{k_i = 1}^{n(M-E_i)} \sum_{k'_{i'} = 1}^{n(M-E_{i'})} 
    \beta^k_{k_i i} \beta^{k*}_{k'_{i'} i'} 
    \ket{\psi_{k_i}(M-E_i)} \ket{\phi_{a+i}} 
    \bra{\psi_{k'_{i'}}(M-E_{i'})} \bra{\phi_{a+i'}}.
\label{eq:emission-mixed}
\end{equation}

The result of the semiclassical calculation implies that
\begin{equation}
  \frac{1}{n(M)} \sum_{k=1}^{n(M)} \beta^k_{k_i i} \beta^{k*}_{k'_{i'} i'} 
  \approx \frac{1}{\bar{Z}} g_i e^{-\frac{E_i}{T_{\rm H}}} 
    \delta_{k_i k'_{i'}} \delta_{ii'},
\label{eq:Hawking-therm}
\end{equation}
where $\bar{Z} = \sum_i n(M-E_i) g_i e^{-E_i/T_{\rm H}}$, and 
$g_i$ is the gray-body factor calculable in the semiclassical 
analysis~\cite{Page:1976df}.  The evolution in Eq.~(\ref{eq:emission-mixed}) 
is then approximated by
\begin{equation}
  \rho(M) \rightarrow 
    \frac{1}{\bar{Z}} \sum_i 
    n(M-E_i) g_i e^{-\frac{E_i}{T_{\rm H}}} \rho_{{\rm fin},i} 
  \approx \frac{1}{Z} \sum_i 
    g_i e^{-\frac{E_i}{T_{\rm H}}} \rho_{{\rm fin},i},
\label{eq:emission-mixed-2}
\end{equation}
where $Z = \sum_i g_i e^{-E_i/T_{\rm H}}$, and we have ignored the $i$ 
dependence of $n(M-E_i)$ in the last expression.  The density matrix
\begin{equation}
  \rho_{{\rm fin},i} = \frac{1}{n(M-E_i)} \sum_{k_i = 1}^{n(M-E_i)} 
    \ket{\psi_{k_i}(M-E_i)} \ket{\phi_{a+i}} 
    \bra{\psi_{k_i}(M-E_i)} \bra{\phi_{a+i}},
\label{eq:rho_fin-i}
\end{equation}
represents the coarse-grained state in which the newly created Hawking 
quanta are in state $i$, so that Eq.~(\ref{eq:emission-mixed-2}) 
corresponds to the well-known result in semiclassical gravity.  Again, 
the final expression in Eq.~(\ref{eq:emission-mixed-2}) does not depend 
on microstates of the black hole, despite the fact that the elementary 
process in Eq.~(\ref{eq:emission-gen}) is unitary, so that the coefficients 
$\beta^k_{k_i i}$ depend on $k$.  This elucidates why the semiclassical 
calculation sees apparent violation of unitarity in the Hawking emission 
process---it deals with the mixed state, Eq.~(\ref{eq:rho_M-2}), from 
the beginning.

\subsection{Interior spacetime}
\label{subsec:interior}

The fact that the semiclassical approximation is built on the maximally 
mixed state, Eq.~(\ref{eq:rho_M}), implies that it cannot probe the index 
$k$ labeling the microstates for the spacetime.  In particular, a structure 
seen in semiclassical gravity is either factored from $k$ or arises 
as a result of taking the maximal mixture.  Here we elucidate how this 
picture works for the emergence of the interior spacetime of a black hole.

Consider a black hole formed by a gravitational collapse.  The spacetime 
region relevant for a distant description is then the region outside the 
event horizon, which we identify as the Schwarzschild horizon ignoring 
a small difference (inessential here) between the two:\ $r > 2 M l_{\rm P}^2 
\equiv R_{\rm S}$.  We may split the degrees of freedom in this region 
into two classes:\ those in $R_{\rm S} < r \leq r_{\rm s}$ and in 
$r > r_{\rm s}$.  The former is the stretched horizon degrees of freedom, 
which are intrinsically quantum gravitational, while the latter is well 
described by a field theory at low energies.  Now, consider Hilbert space 
${\cal H}_{\tilde{\psi}_k}$ spanned by $\ket{\psi_k(M)}$ and all the 
excited states of it, $\ket{\tilde{\psi}_{kn}(M)}$, for fixed $k$. 
Assuming that the number of excitations is sufficiently small that 
Eq.~(\ref{eq:S_excit}) is satisfied, a basis vector of this Hilbert 
space is effectively specified by the state of the stretched horizon, 
labeled by $\tilde{\imath}$, and the number of excitations $n_\sigma$ 
in each field theory mode $\sigma$ in the zone region:
\begin{equation}
  \ket{\tilde{\psi}_{k;\tilde{\imath} \{ n_\sigma \}}(M)} 
  = \ket{\tilde{\imath}; k} \ket{\{ n_\sigma \}; k}.
\label{eq:basis}
\end{equation}
Here, we assume that the field theory modes $\sigma$ are defined using 
the Schwarzschild time $t$ at a sufficiently late time when the geometry 
in the region $r > R_{\rm S}$ is well approximated by the Schwarzschild 
spacetime.

Note that in Eq.~(\ref{eq:basis}), we have kept the index $k$ in the 
stretched horizon and zone states to remind ourselves that they represent 
excitations on the black hole vacuum state $\ket{\psi_k(M)}$.  This, 
however, does not mean that the states $\ket{\tilde{\imath}; k}$ and 
$\ket{\tilde{\imath}; k'}$ for $k \neq k'$, or $\ket{\{ n_\sigma \}; k}$ 
and $\ket{\{ n_\sigma \}; k'}$ for $k \neq k'$, are all independent. 
In fact, to see the structure of the independence of these states, 
it is better to write Eq.~(\ref{eq:basis}) in a form analogous to 
Eq.~(\ref{eq:separate'}):
\begin{equation}
  \ket{\tilde{\psi}_{k;\tilde{\imath} \{ n_\sigma \}}(M)} 
  = \ket{\tilde{\imath}, g} \ket{\{ n_\sigma \}, h},
\label{eq:basis-2}
\end{equation}
where $k = \{ g, h \}$ with $g = 1,\cdots,G$ and $h = 1,\cdots,H$.  The 
states $\ket{\tilde{\imath}, g}$ and $\ket{\{ n_\sigma \}, h}$ may then 
be viewed as all independent:
\begin{equation}
  \inner{\tilde{\imath}, g}{\tilde{\imath}, g'} = \delta_{gg'},
\qquad
  \inner{\{ n_\sigma \}, h}{\{ n_\sigma \}, h'} = \delta_{hh'},
\label{eq:basis-3}
\end{equation}
with
\begin{equation}
  GH = n(M).
\label{eq:basis-4}
\end{equation}
We expect that the dimensions of the Hilbert space factors for the 
stretched horizon and zone states are roughly comparable, $\ln G 
\approx \ln H \approx O(M^2 l_{\rm P}^2)$; their precise ratio depends 
on how we divide the degrees of freedom between the stretched horizon 
and the zone for the modes around $r \approx r_{\rm s}$.  In the rest 
of the paper we adopt the notation in Eq.~(\ref{eq:basis}), which 
makes the connection to the vacuum states clearer.  It should, however, 
be understood that the index $k$ represents the information that is 
{\it shared} between (and not possessed by both) the stretched horizon 
and zone states, in the sense of Eqs.~(\ref{eq:basis-2},~\ref{eq:basis-4}).

Now, the analysis in semiclassical gravity tells us that to describe the 
interior of a black hole, we need quantum field theory modes corresponding 
to the {\it second} exterior region of a maximally extended---or 
two-sided---black hole, in addition to the ``original'' exterior modes 
$\sigma$~\cite{Unruh:1976db,Israel:1976ur}.  Where do such modes come 
from?  We assume that the required modes arise from excitations of the 
stretched horizon degrees of freedom.  In particular, the intrinsically 
quantum gravitational dynamics at the stretched horizon organize these 
modes such that they ``mirror'' the structure of the near exterior modes, 
which can be interpreted as being located in the region outside the 
stretched horizon of the second exterior region~\cite{Nomura:2013lia}. 
In order to reproduce the relevant interior spacetime region, the 
stretched horizon has to mirror the modes in the region $r^*_{\rm s} 
< r^* < 0$ (corresponding to the causal past of a $t=0$ point on 
singularities, assuming the reflection symmetry between the two exterior 
regions), but it need not do more; in particular, we expect that the 
stretched horizon does not produce modes mirroring the far exterior modes.

General near horizon black hole states built on a vacuum state 
$\ket{\psi_k(M)}$ may then be written as
\begin{equation}
  \ket{\tilde{\psi}_k(M)} = \sum_{\{ \tilde{n}_\sigma \}, \{ n_\sigma \}} 
    d_{\{ \tilde{n}_\sigma \} \{ n_\sigma \}}\, 
    \ket{\{ \tilde{n}_\sigma \}; k} \ket{\{ n_\sigma \}; k}.
\label{eq:near-horizon}
\end{equation}
The stretched horizon states are now labeled by the set of occupation 
numbers, $\tilde{n}_\sigma$, for all the mirror modes, which we have 
labeled using the same symbol $\sigma$ as the original modes.%
\footnote{In a distant description, a stretched horizon mode specified 
 by $\sigma$, being intrinsically quantum gravitational degrees of freedom, 
 need not have the same localization property in the angular directions 
 as the corresponding mode in the near exterior region.}
We can now define mode operators $\tilde{b}_\sigma^{(k)}$ and 
$\tilde{b}_\sigma^{(k)\dagger}$ acting on the states for each $k$ as
\begin{align}
  \tilde{b}_\sigma^{(k)} \ket{\{ \tilde{n}_{\sigma'} \}; k'} 
    \ket{\{ n_{\sigma''} \}; k'} 
  &= \delta_{kk'} \sqrt{\tilde{n}_\sigma}\, 
      \ket{\{ \tilde{n}_{\sigma'} - \delta_{\sigma\sigma'} \}; k} 
      \ket{\{ n_{\sigma''} \}; k},
\label{eq:tilde-b}\\
  \tilde{b}_\sigma^{(k)\dagger} \ket{\{ \tilde{n}_{\sigma'} \}; k'} 
    \ket{\{ n_{\sigma''} \}; k'} 
  &= \delta_{kk'} \sqrt{\tilde{n}_\sigma + 1}\, 
      \ket{\{ \tilde{n}_{\sigma'} + \delta_{\sigma\sigma'} \}; k} 
      \ket{\{ n_{\sigma''} \}; k},
\label{eq:tilde-b-dag}
\end{align}
analogously to the annihilation-creation operators $b_{\sigma}^{(k)}$ 
and $b_\sigma^{(k)\dagger}$ for the near exterior modes:
\begin{align}
  b_{\sigma}^{(k)} \ket{\{ \tilde{n}_{\sigma''} \}; k'} 
    \ket{\{ n_{\sigma'} \}; k'} 
  &= \delta_{kk'} \sqrt{n_\sigma}\, 
      \ket{\{ \tilde{n}_{\sigma''} \}; k} 
      \ket{\{ n_{\sigma'} - \delta_{\sigma\sigma'} \}; k},
\label{eq:b}\\
  b_\sigma^{(k)\dagger} \ket{\{ \tilde{n}_{\sigma''} \}; k'} 
    \ket{\{ n_{\sigma'} \}; k'} 
  &= \delta_{kk'} \sqrt{n_\sigma + 1}\, 
      \ket{\{ \tilde{n}_{\sigma''} \}; k} 
      \ket{\{ n_{\sigma'} + \delta_{\sigma\sigma'} \}; k}.
\label{eq:b-dag}
\end{align}
These operators then satisfy the commutation relations
\begin{align}
  & [\tilde{b}_\sigma^{(k)}, \tilde{b}_{\sigma'}^{(k')\dagger}] 
  = \delta_{\sigma\sigma'} \delta_{kk'} P_k,
\qquad
  [\tilde{b}_\sigma^{(k)}, \tilde{b}_{\sigma'}^{(k')}] 
  = [\tilde{b}_\sigma^{(k)\dagger}, \tilde{b}_{\sigma'}^{(k')\dagger}] = 0,
\label{eq:tilde-b_k-comm}\\
  & [b_\sigma^{(k)}, b_{\sigma'}^{(k')\dagger}] 
  = \delta_{\sigma\sigma'} \delta_{kk'} P_k,
\qquad
  [b_\sigma^{(k)}, b_{\sigma'}^{(k')}] 
  = [b_\sigma^{(k)\dagger}, b_{\sigma'}^{(k')\dagger}] = 0,
\label{eq:b_k-comm}
\end{align}
which imply that we can interpret $\tilde{b}_\sigma^{(k)}$ and 
$\tilde{b}_\sigma^{(k)\dagger}$ as the annihilation-creation operators for 
``mirror quanta'' arising as collective excitation modes of the stretched 
horizon degrees of freedom.  Here, $P_k$ is the projection operator 
on ${\cal H}_{\tilde{\psi}_k}$, i.e.\ $P_k \ket{\tilde{\psi}_{k'}(M)} 
= \delta_{kk'} \ket{\tilde{\psi}_k(M)}$, which appears because 
$\tilde{b}_\sigma^{(k)}$ and $b_\sigma^{(k)}$ involve projection 
on ${\cal H}_{\tilde{\psi}_k}$, i.e.\ $\tilde{b}_\sigma^{(k)} = 
\tilde{b}_\sigma^{(k)} P_k$ and $b_\sigma^{(k)} = b_\sigma^{(k)} P_k$. 
Note that this is not a local operator in the usual sense of field 
theory, since the information about $k$ is spread over the stretched 
horizon as well as the zone regions.%
\footnote{We consider that the states obtained by acting operators 
 $b_{\sigma}^{(k)}$, $b_\sigma^{(k)\dagger}$, $\tilde{b}_\sigma^{(k)}$, 
 and $\tilde{b}_\sigma^{(k)\dagger}$ on a black hole vacuum state are 
 not all physically realized states, so that the actual physical Hilbert 
 space is smaller than the Fock space implied by the construction here. 
 In particular, the dimension of the physical Hilbert space for the 
 states in which there are negative energy excitations (as defined 
 in the asymptotic region) must be smaller than that of the naive Fock 
 space described here, implying that the entropy $S_I$ associated with 
 a negative energy excitation $I$ is negative in Eq.~(\ref{eq:S_excit}). 
 For more discussions on this point, see Section~\ref{subsec:acceleration}.}

The construction of the operators relevant for describing the black hole 
spacetime can now go as in the semiclassical analysis~\cite{Unruh:1976db}. 
Specifically, we can split the modes $\sigma$ into two classes:\ outgoing, 
$\sigma^+$, and ingoing, $\sigma^-$.  For the outgoing modes, we introduce 
the mode operators associated with the Killing vector on the past horizon, 
$\partial/\partial U$ where $U = -M l_{\rm P}^2 e^{(r^*-t)/4 M l_{\rm P}^2}$, 
in the analytically extended black hole background:
\begin{equation}
  a_\xi^{(k)} 
  = \sum_{\sigma^+} \bigl( \alpha_{\xi\sigma^+} b_{\sigma^+}^{(k)} 
    + \gamma_{\xi\sigma^+} b_{\sigma^+}^{(k)\dagger} 
    + \zeta_{\xi\sigma^+} \tilde{b}_{\sigma^+}^{(k)} 
    + \eta_{\xi\sigma^+} \tilde{b}_{\sigma^+}^{(k)\dagger} \bigr),
\label{eq:a_xi}
\end{equation}
where $\xi$ represents the quantum numbers labeling the modes, and 
$\alpha_{\xi\sigma^+}$, $\gamma_{\xi\sigma^+}$, $\zeta_{\xi\sigma^+}$, 
and $\eta_{\xi\sigma^+}$ are the Bogoliubov coefficients, calculable 
using the standard quantum field theory method.  These operators satisfy 
the commutation relations for annihilation-creation operators
\begin{equation}
  [a_\xi^{(k)}, a_{\xi'}^{(k')\dagger}] 
  = \delta_{\xi\xi'} \delta_{kk'} P_k,
\qquad
  [a_\xi^{(k)}, a_{\xi'}^{(k')}] 
  = [a_\xi^{(k)\dagger}, a_{\xi'}^{(k')\dagger}] = 0.
\label{eq:a_k-comm}
\end{equation}
The near horizon black hole vacuum state $\ket{\psi_k(M)}$ is then well 
approximated by the condition
\begin{equation}
  \forall \xi, \sigma^- \quad 
  a_\xi^{(k)} \ket{\psi_k(M)} = b_{\sigma^-}^{(k)} \ket{\psi_k(M)} 
  = \tilde{b}_{\sigma^-}^{(k)} \ket{\psi_k(M)} = 0.
\label{eq:inf_vac_k}
\end{equation}
Ignoring any excitations, this state takes approximately the thermofield 
double form:
\begin{equation}
  \ket{\psi_k(M)} = 
    \frac{1}{\sqrt{Z}} \sum_{\{ n_{\sigma^+} \}} 
    e^{-\frac{E_{\{ n_{\sigma^+} \}}}{2 T_{\rm H}}} 
    \ket{\{ \tilde{n}_{\sigma^+} = n_{\sigma^+} \}; k} 
    \ket{\{ n_{\sigma^+} \}; k};
\qquad
  Z = \sum_{\{ n_{\sigma^+} \}} 
    e^{-\frac{E_{\{ n_{\sigma^+} \}}}{T_{\rm H}}},
\label{eq:psi_k}
\end{equation}
where $E_{\{ n_{\sigma^+} \}}$ is the energy of the state 
$\ket{\{ n_{\sigma^+} \}; k}$ as measured in the asymptotic 
region, and the occupation numbers for the ingoing modes in 
$\ket{\{ \tilde{n}_{\sigma^+} = n_{\sigma^+} \}; k}$ and 
$\ket{\{ n_{\sigma^+} \}; k}$ are zero.

In general, the microscopic Hamiltonian in quantum gravity may depend in a 
complicated way on the sets of operators $a_\xi^{(k)}$, $b_{\sigma^-}^{(k)}$, 
and $\tilde{b}_{\sigma^-}^{(k)}$ as well as other operators, including the 
ones that act nontrivially on the index $k$:
\begin{equation}
  H(M) = H_{\rm QG}\bigl( \bigl\{ a_\xi^{(k)} \bigr\}, 
    \bigl\{ b_{\sigma^-}^{(k)} \bigr\}, 
    \bigl\{ \tilde{b}_{\sigma^-}^{(k)} \bigr\}, \cdots; M \bigr),
\label{eq:H_QG}
\end{equation}
even if we restrict our considerations to the states having spacetime with 
a black hole at a fixed location of mass between $M$ and $M + \delta M$. 
(If we consider processes in which $M$ varies, the total Hamiltonian 
must be taken as the sum of the $H(M)$'s with different $M$'s, with terms 
allowing hopping between different $M$'s added.  Of course, the segmentation 
of the black hole mass into different discrete $M$'s with some widths 
is completely artificial, and thus can be done in any arbitrary way.)
On the other hand, the semiclassical approximation is built on the 
maximally mixed vacuum state, Eq.~(\ref{eq:rho_M}).  What is the precise 
meaning of this statement?

Consider a general semiclassical state
\begin{equation}
  \tilde{\rho}(M) 
  = f\bigl( a_\xi^\dagger, b_{\sigma^-}^\dagger, 
      \tilde{b}_{\sigma^-}^\dagger \bigr)\, \rho(M)\, 
    f\bigl( a_\xi^\dagger, b_{\sigma^-}^\dagger, 
      \tilde{b}_{\sigma^-}^\dagger \bigr)^\dagger,
\label{eq:tilde-rho_M}
\end{equation}
where $\rho(M)$ is the semiclassical vacuum state in Eq.~(\ref{eq:rho_M}), 
and $f$ represents the excitation.  Here, $a_\xi$, $b_{\sigma^-}$, and 
$\tilde{b}_{\sigma^-}$ are (semiclassical) quantum field theory operators 
defined by
\begin{equation}
  a_\xi = \sum_k a_\xi^{(k)},
\qquad
  b_{\sigma^-} = \sum_k b_{\sigma^-}^{(k)},
\qquad
  \tilde{b}_{\sigma^-} = \sum_k \tilde{b}_{\sigma^-}^{(k)},
\label{eq:modes}
\end{equation}
which satisfy the commutation relations
\begin{equation}
  [a_\xi, a_{\xi'}^\dagger] = \delta_{\xi\xi'} \sum_k P_k,
\qquad
  [b_{\sigma^-}, b_{\sigma^{-\prime}}^\dagger] 
  = [\tilde{b}_{\sigma^-}, \tilde{b}_{\sigma^{-\prime}}^\dagger] 
  = \delta_{\sigma^- \sigma^{-\prime}} \sum_k P_k,
\qquad
  \mbox{others } = 0.
\label{eq:a-comm}
\end{equation}
Note that the operator $\sum_k P_k$ becomes unity when acting on any 
normalized (pure or mixed) state in $\bigoplus_k {\cal H}_{\tilde{\psi}_k}$, 
so that Eq.~(\ref{eq:a-comm}) indeed represents the standard commutation 
relations for the annihilation-creation operators.

The semiclassical approximation asserts that the evolution of the state 
in Eq.~(\ref{eq:tilde-rho_M}), which at the microscopic level is generated 
by the Hamiltonian in Eq.~(\ref{eq:H_QG}), is well approximated by the 
evolution caused by the standard quantum field theory Hamiltonian:
\begin{equation}
  \tilde{\rho}(M) \rightarrow
    e^{-iH(M)t}\, \tilde{\rho}(M)\, e^{iH(M)t},
\qquad
  H(M) \approx 
    H_{\rm QFT}\bigl( a_\xi, b_{\sigma^-}, \tilde{b}_{\sigma^-}; M \bigr),
\label{eq:evolution}
\end{equation}
where the dependence of $H_{\rm QFT}$ on the creation, as well 
as annihilation, operators is implied.  Note that here we have 
used the $a_\xi$ operators to describe the outgoing modes in the 
semiclassical field theory, but we may instead use $b_{\sigma^+}$ 
and $\tilde{b}_{\sigma^+}$ operators defined analogously to 
Eq.~(\ref{eq:modes}):
\begin{equation}
  b_{\sigma^+} = \sum_k b_{\sigma^+}^{(k)},
\qquad
  \tilde{b}_{\sigma^+} = \sum_k \tilde{b}_{\sigma^+}^{(k)}.
\label{eq:modes-2}
\end{equation}
These two descriptions correspond, respectively, to seeing the system from 
an infalling and static (or distant) observer's viewpoints.%
\footnote{The $\xi$ modes correspond to the modes defined in the past null 
 infinity using the Minkowski time if we treat the collapsing geometry 
 faithfully (without invoking the eternal black hole approximation at 
 late times) ignoring the trans-Planckian nature of the encounters of 
 the modes with the collapsing matter.}

The expression in Eq.~(\ref{eq:evolution}) makes it clear that 
semiclassical gravity cannot describe detailed microscopic physics 
associated with the index $k$.  In fact, at this level of approximation, 
the Hamiltonian can be written as
\begin{equation}
  H(M) \approx 
    H_{\rm QFT}\bigl( a_\xi, b_{\sigma^-}, \tilde{b}_{\sigma^-}; M \bigr) 
  = \sum_k H_{\rm QFT}\bigl( a_\xi^{(k)}, b_{\sigma^-}^{(k)}, 
      \tilde{b}_{\sigma^-}^{(k)}; M \bigr),
\label{eq:H_QFT-2}
\end{equation}
since $a_\xi^{(k)}, b_{\sigma^-}^{(k)}$, and $\tilde{b}_{\sigma^-}^{(k)}$ 
(as well as $b_{\sigma^+}^{(k)}$ and $\tilde{b}_{\sigma^+}^{(k)}$) involve 
projection on ${\cal H}_{\tilde{\psi}_k}$.  This implies that the dynamics 
involving (only) excitations on a fixed black hole background is decomposed 
into $n(M) = e^{4\pi M^2 l_{\rm P}^2}$ decoupled, identical copies.  Since 
this statement also applies to any fixed spatial geometry arising for a 
sufficiently short time interval (with the appropriate replacement of the 
number of microstates), we may simply drop the index $k$ in describing the 
dynamics of the excitations, and write the evolution of the system as
\begin{equation}
  \ket{\tilde{\Psi}} \rightarrow e^{-iH_{\rm QFT} t}\, \ket{\tilde{\Psi}}.
\label{eq:QFT}
\end{equation}
Here, $\ket{\tilde{\Psi}}$ is a field theoretical state of the system 
built on a field theoretical vacuum $\ket{\Psi}$, e.g., $\ket{\tilde{\Psi}} 
= f( a_\xi^\dagger, b_{\sigma^-}^\dagger, \tilde{b}_{\sigma^-}^\dagger) 
\ket{\Psi}$ with $f$ in Eq.~(\ref{eq:tilde-rho_M}).  This is what the 
semiclassical field theory is!  It provides a unitary description of 
physics {\it unless} we consider processes involving the microscopic 
spacetime index $k$ such as the Hawking emission process.

\subsection{UV/IR correspondence and relation to the entanglement entropy}
\label{subsec:duality}

Before concluding this section, we would like to highlight a key assumption 
adopted in Section~\ref{subsec:interior} as well as its implications. 
The assumption may be called the {\it stretched horizon/second exterior 
(SH/SE) correspondence}:
\begin{itemize}
\item[]
The stretched horizon degrees of freedom representing excitations on 
a black hole vacuum state $\ket{\psi_k(M)}$ can be organized into the 
modes that are interpreted as being located outside the stretched horizon 
of the second (near) exterior region of the corresponding extended, or 
two-sided, black hole of mass $M$.
\end{itemize}
Under this correspondence, the ``trans-Planckian''---or stringy---excitations 
in the region $R_{\rm S} < r \leq r_{\rm s}$ in the ``original'' one-sided 
picture are mapped into low energy field theory excitations outside the 
stretched horizon of the second exterior region in the corresponding 
two-sided picture.  This implies that there are no (or few) excitations 
at a distance below the quantum gravitational lengthscale in the latter 
picture (at least around the bifurcation surface).  We may therefore 
colloquially summarize this correspondence as:\ the full string theory 
excitations defined in the region outside the mathematical Schwarzschild 
horizon, $r > R_{\rm S}$, correspond to field theory excitations defined 
in the extended, two-sided black hole spacetime.

While we have motivated the correspondence by the physical picture in 
semiclassical gravity, it would be desirable to derive or motivate 
it more directly in the fundamental theory of quantum gravity.  In 
fact, we view this as an interesting opportunity.  The description in 
Section~\ref{subsec:interior} requires specific properties for the dynamics 
of the stretched horizon (the UV degrees of freedom above the quantum 
gravitational scale), reminiscent of UV/IR relations seen in other 
settings in string theory (such as AdS/CFT and $T$ duality).  This 
therefore provides an explicit suggestion on what properties the fundamental 
theory of quantum gravity must have in order to reproduce the predictions 
of general relativity in the appropriate classical limit, in situations 
in which spacetime is far from simple Minkowski or anti-de~Sitter space.

The picture described here also provides a simple interpretation of possible 
relations between the black hole and entanglement entropies contemplated 
long ago in Ref.~\cite{Sorkin}.  As we have discussed, the relevant 
spacetime region in a distant description of a black hole is the outside 
of the mathematical horizon $r > R_{\rm S}$, namely the near side of the 
Schwarzschild horizon as viewed from the ``origin of the reference frame'' 
located outside the horizon.  The number of independent quantum states 
in full quantum gravity describing this region---more precisely the near 
horizon part of it---is given by $n(M) = e^{{\cal A}/4 l_{\rm P}^2}$, 
where ${\cal A}$ is the horizon area.%
\footnote{This statement may be generalized to a more concrete statement 
 if we include the ``boundary'' of the ambient space as viewed from the 
 origin of the reference frame $p_0$:\ the number of independent quantum 
 states representing the region enclosed by the black hole horizon 
 and the boundary (as viewed from $p_0$) is given by the exponential 
 of the sum of the horizon and boundary areas in units of $4l_{\rm P}^2$. 
 (In Minkowski space, the boundary may be located at infinite distances 
 away and may have an infinite area.)  For further discussions, see 
 Section~\ref{sec:discuss} (and Ref.~\cite{Nomura:2013nya}).}
How can we calculate this number without invoking the thermodynamic 
argument as was originally done by Bekenstein and Hawking?

One way to do so is the following.  We {\it fictitiously} extend the 
spacetime beyond the horizon $r = R_{\rm S}$ by considering another 
``mirror'' region parameterized by $r' > R_{\rm S}$ sewn to the original 
region at $r = r' = R_{\rm S}$.  (Note that the mirror region, $r' 
> R_{\rm S}$, in this discussion is really fictitious, since we are 
describing the system from a distant viewpoint using the fundamental 
theory, so that the degrees of freedom in the original region, 
$r > R_{\rm S}$, already contain the stretched horizon degrees 
of freedom, which are sufficient to construct the interior of the 
black hole.)  We may then consider a state in which the degrees of 
freedom in the two regions, $r > R_{\rm S}$ and $r' > R_{\rm S}$, 
are (nearly) maximally entangled.  The number of independent states 
available in one region can then be obtained as the exponential of 
the entanglement entropy between the two regions.

For black hole spacetime, we may consider the intersection of the two 
regions, $r = r' = R_{\rm S}$, to be a bifurcation surface, and the mirror 
region to be the other side of that surface on an equal-time hypersurface 
passing through it.  Since we are interested in the number of microstates 
in full quantum gravity, this construction must be done in the full theory 
of quantum gravity, implying that the mirror region, $r' > R_{\rm S}$, 
also ``mirrors'' the effective second exterior region of the extended 
black hole, arising from the stretched horizon modes in the original 
region.  Now, we expect that the local inertial vacuum state around 
$r \sim R_{\rm S}$ in this ``doubled spacetime'' involves near maximal 
entanglement between the states in the two regions, $r > R_{\rm S}$ and 
$r' > R_{\rm S}$, when viewed from a static/distant observer because of 
a large relative acceleration between the two reference frames.  We may 
therefore obtain the black hole entropy by calculating the entanglement 
entropy between the two regions for the local inertial vacuum state in 
the full theory of quantum gravity.  This corresponds to the calculation 
performed in Ref.~\cite{Susskind:1994sm} in string theory, which indeed 
found that the entanglement entropy is ${\cal A}/4 l_{\rm P}^2$ at 
the leading order in $l_{\rm P}^2/{\cal A}$.

Do we really need string theory to calculate the entanglement entropy? 
In low energy quantum field theories, the result of the calculation will 
be divergent, which must be made finite by counterterms.  The freedom 
in adjusting these counterterms represents a possible variation of UV 
theories that lead to consistent quantum field theories at low energies. 
At the leading order in $l_{\rm P}^2/{\cal A}$, however, no such freedom 
is left after the Planck length, $l_{\rm P}$, is renormalized to a finite 
value, and the calculation in Ref.~\cite{Cooperman:2013iqr} indeed finds 
that the entanglement entropy after this renormalization is finite and 
takes an unambiguous value of ${\cal A}/4 l_{\rm P}^2$.  (To be precise, 
Ref.~\cite{Cooperman:2013iqr} showed this only for certain limited 
cases, but we may expect it applies more generally.)  This implies 
that any consistent UV theory leads to the entanglement entropy between 
the two regions of ${\cal A}/4 l_{\rm P}^2$ in the local inertial vacuum 
state.  As argued above, this measures the number of independent quantum 
states available in one of the regions in the full theory of quantum 
gravity.  In the context of a black hole described in a distant reference 
frame, these states comprise all of the states in the near horizon 
region, since the mirror region $r' > R_{\rm S}$ (again, not to be 
confused with the second exterior region in the extended black hole 
approximation) does not really exist.

\section{Nature of Microscopic Dynamics}
\label{sec:micro}

We now discuss the nature of microscopic dynamics associated with Hawking 
emission and measuring the black hole's thermal atmosphere by a physical 
detector (the mining process).  A complete treatment of this issue 
requires the fundamental theory of quantum gravity.  Our focus here is 
how these dynamics manifest themselves in the semiclassical approximation, 
in particular what physical conclusions we may draw using semiclassical 
analyses.

We argue that a measurement of the black hole's atmosphere by a static 
detector consists of two effects:\ the effect caused by acceleration 
of the detector (the Unruh effect) and that by nonzero curvature of 
spacetime.  The two combined make the detector respond as if it is 
immersed in the thermal bath at a blueshifted local Hawking temperature. 
If the detector is located in a region away from the black hole, then 
the former acceleration induced effect becomes negligible, and its 
response is dominated by the latter curvature induced effect.  It is 
this latter effect that represents the Hawking radiation emitted to 
the asymptotic region.

We assert that the microscopic details associated with the information 
transfer are in general not visible in semiclassical gravity, although 
a part of the backreaction of the acceleration induced effect can be 
described in semiclassical field theory.  In particular, this implies 
that at the level of semiclassical gravity, the curvature induced effect 
is described only as ``physics of a vacuum,'' which arises from an 
intrinsic quantum mechanical ambiguity of local vacuum energy density 
(or particle numbers) of order $1/(M l_{\rm P}^2)^4$ that cannot be 
resolved in semiclassical field theory.  One must therefore be careful 
in applying intuition from semiclassical field theory in analyzing 
the microscopic processes underlying Hawking radiation.

\subsection{Response of a static detector}
\label{subsec:response}

Consider a physical detector located at a fixed Schwarzschild radial 
coordinate $r = r_{\rm d}$.  According to the semiclassical calculation, 
the detector responds as if it is immersed in the thermal bath of 
temperature $T(r_{\rm d})$ in Eq.~(\ref{eq:T_rad}).  What causes 
this phenomenon?

First of all, since the detector at a fixed coordinate point $r = r_{\rm d}$ 
is accelerated with respect to local inertial frames, the effect in 
Ref.~\cite{Unruh:1976db}---the Unruh effect---makes the detector react 
accordingly.  The magnitude of the proper acceleration of the detector 
is given by
\begin{equation}
  a(r_{\rm d}) = \frac{1}{\sqrt{1-\frac{2 M l_{\rm P}^2}{r_{\rm d}}}} 
    \frac{M l_{\rm P}^2}{r_{\rm d}^2}.
\label{eq:accel}
\end{equation}
If this were the only effect, then the detector would see a thermal bath 
of temperature
\begin{equation}
  T_{\rm U}(r_{\rm d}) = \frac{a(r_{\rm d})}{2\pi} 
  = \frac{M l_{\rm P}^2}{2\pi r_{\rm d}^2 
    \sqrt{1-\frac{2Ml_{\rm P}^2}{r_{\rm d}}}},
\label{eq:T_U}
\end{equation}
which does not agree with $T(r_{\rm d})$ except when the detector is 
located at the horizon, $r_{\rm d} \rightarrow 2 M l_{\rm P}^2$.  In 
particular, for $r_{\rm d} \rightarrow \infty$ the effect from acceleration 
disappears, $T_{\rm U}(r_{\rm d}) \rightarrow 0$, so that it cannot 
be responsible for Hawking radiation measured at the asymptotic 
infinity.  What is the remaining effect making up the difference 
between $T_{\rm U}(r_{\rm d})$ and $T(r_{\rm d})$?

To look for that effect, we can make the following heuristic argument. 
Consider a detector placed somewhere in the zone.  If there were the 
only acceleration effect, the local energy density at the detector 
location (as defined in the asymptotic region) would be
\begin{equation}
  \rho_{\rm U}(r_{\rm d}) 
  = \frac{c}{1-\frac{2 M l_{\rm P}^2}{r_{\rm d}}} 
    \left( \frac{M l_{\rm P}^2}{2\pi r_{\rm d}^2} \right)^4,
\label{eq:rho_U}
\end{equation}
where $c = \pi^2 g_*/30$ is a numerical coefficient with $g_*$ being the 
effective number of relativistic degrees of freedom, and the quantity in 
the parentheses is the Unruh temperature in Eq.~(\ref{eq:T_U}) corrected 
by the redshift factor $\sqrt{1-2Ml_{\rm P}^2/r_{\rm d}}$.  On the other 
hand, the semiclassical analysis tells us that this quantity must be
\begin{equation}
  \rho(r_{\rm d}) 
  = \frac{c\, T_{\rm H}^4}{1-\frac{2 M l_{\rm P}^2}{r_{\rm d}}} 
  = \frac{c}{1-\frac{2 M l_{\rm P}^2}{r_{\rm d}}} 
    \left( \frac{1}{8\pi M l_{\rm P}^2} \right)^4,
\label{eq:rho}
\end{equation}
consistent with the blueshifted Hawking temperature.  The difference is 
given by
\begin{equation}
  \rho(r_{\rm d}) - \rho_{\rm U}(r_{\rm d}) 
  = c \left( \frac{1}{8\pi M l_{\rm P}^2} \right)^4 
    f\Bigl(\frac{r_{\rm d}}{2 M l_{\rm P}^2}\Bigr);
\qquad
  f(x) = \frac{1 - \frac{1}{x^8}}{1 - \frac{1}{x}}.
\label{eq:diff-rho}
\end{equation}
Since $f(x)$, defined for $x > 1$, is a monotonically decreasing function 
with $f(x \rightarrow 1) = 8$ and $f(x \gg 1) \rightarrow 1$, we find that 
the effect we seek is giving a contribution
\begin{equation}
  \rho(r) - \rho_{\rm U}(r) 
  \approx O\biggl( \frac{1}{(8\pi M l_{\rm P}^2)^4} \biggr),
\label{eq:remaining}
\end{equation}
throughout the entire zone region.

The local energy density of the amount in Eq.~(\ref{eq:remaining}) 
is precisely what we would expect to arise from the intrinsic quantum 
mechanical ambiguity of defining particle states in semiclassical 
gravity.  Since the curvature lengthscale around the black hole is 
of order $M l_{\rm P}^2$, the concept of particles is ambiguous 
beyond this lengthscale.  This makes particle states with $\omega 
\simlt 1/M l_{\rm P}^2$ ambiguous, leading to an uncertainty of the 
local vacuum energy density of order $1/(M l_{\rm P}^2)^4$.  Here, 
$\omega$ is the frequency as defined in the asymptotic region.  We 
assume that this curvature induced effect is responsible for the 
difference in Eq.~(\ref{eq:remaining}), i.e.\ the difference between 
$T_{\rm U}(r_{\rm d})$ and $T(r_{\rm d})$.  In particular, this effect 
is solely responsible for the response of a detector located in the 
far asymptotic region, i.e.\ the original Hawking radiation effect 
calculated using the in-out formalism~\cite{Hawking:1974sw}.  Unlike 
the acceleration induced effect, this effect cannot be eliminated 
by going to a local inertial frame.

The separation of the origin of the detector response into the two effects 
as described above is useful in discussing what aspects of the underlying 
process are visible in semiclassical gravity.  At the microscopic level, 
both these effects occur according to the full Hamiltonian evolution in 
quantum gravity, and all the microscopic details associated with them can 
be calculated, at least, in principle.  The same, however, is not true in 
the semiclassical approximation.  We now see to what extent the details 
of the underlying dynamics can be captured in semiclassical field theory 
for each of these two effects.

\subsection{Effect from acceleration}
\label{subsec:acceleration}

Imagine that a physical detector is held near the horizon.  The detector 
then responds to a highly blueshifted Hawking temperature.  As can be 
seen from the fact that
\begin{equation}
  \frac{\rho_{\rm U}(r)}{\rho(r) - \rho_{\rm U}(r)} 
  = \frac{1}{\bigl( \frac{r}{2 M l_{\rm P}^2} \bigr)^4 - 1} 
  \gg 1
\quad\mbox{for}\quad
  r \simeq 2 M l_{\rm P}^2,
\label{eq:rho-ratio}  
\end{equation}
this response is caused mostly by the acceleration of the detector with 
respect to local inertial frames.  The dynamics of this effect, including 
the backreaction, was analyzed in Ref.~\cite{Unruh:1983ms} in semiclassical 
field theory, which we may import to study our black hole problem here.

Suppose the detector is coupled to a field $\varphi$.  When described 
in a static reference frame, the backreaction of a detector response 
eliminates a $\varphi$ particle from the thermal bath, interpreted as 
being absorbed by the detector:\ $\ket{\Psi(M)} \rightarrow b_\sigma 
\ket{\Psi(M)}$, where $\ket{\Psi(M)}$ represents the field theory black 
hole vacuum, and an appropriate superposition of $b_\sigma$ is implied. 
In an infalling frame, however, the same process is described as an 
{\it emission} of a $\varphi$ particle by the detector, i.e.\ the 
backreaction of the detector response is the creation of a particle, 
as can be seen from the fact that $b_\sigma$ involves a superposition 
of $a_\xi$ and $a_\xi^\dagger$, and $a_\xi \ket{\Psi(M)} = 0$.  As 
expected, the effect of this backreaction is limited within the causal 
future of the detector response event.  In this region, however, the 
backreaction indeed {\it does} make the system deviate from the vacuum 
state, which is visible in the semiclassical approximation.  In fact, 
the effect can be quite significant if the detector is located very 
close to the horizon.

It is important to understand the physical origin of this backreaction. 
Suppose we describe the detector in an infalling reference frame.  When 
it responds, i.e.\ gets excited, it emits a $\varphi$ particle to the 
environment, so that the (appropriately defined) total energy of the 
system increases.  Where does this energy come from?  In order for the 
detector to measure a $\varphi$ particle, it must be held at $r = r_{\rm d}$ 
for a Schwarzschild time of order $\varDelta t \approx M l_{\rm P}^2$. 
This requires an external force---without it, the detector would fall 
into the horizon within a much shorter time.  It is this force that is 
responsible for the energy needed for the detector response as well as 
the $\varphi$ emission, i.e.\ the backreaction.  Since the vacuum is 
physically disturbed by the external non-gravitational force, it is 
no surprise that there is a backreaction effect visible clearly in the 
semiclassical approximation.

What about the information transfer?  It is known that one can 
accelerate the energy loss rate of a black hole by extracting its 
energy from the atmosphere using a physical apparatus:\ the mining 
process~\cite{Unruh:1982ic}.  The most efficient, and essentially the 
only, way to do this is to thread the horizon with strings that have 
the maximal tension-to-linear-mass-density ratio allowed by the null 
energy condition, which can make the black hole lifetime as short 
as $O(M^2 l_{\rm P}^3)$~\cite{Brown:2012un}.  Now, the fact that the 
energy is extracted by the apparatus means that the information must 
also be extracted, since otherwise the entropy of the black hole 
would be oversaturated beyond the Bekenstein-Hawking value.  How 
can information be transferred in such a process?%
\footnote{Our analysis in the previous version of this paper on this 
 issue contained an error. We thank Joseph Polchinski for a comment 
 that made us identify and correct that error.}

Suppose we mine a black hole with the model ``detector'' considered 
before, around Eqs.~(\ref{eq:probe}) and (\ref{eq:int-mixed}).  This 
detector may be viewed as a toy model for the strings described above. 
The evolution of the combined black hole and detector system then 
occurs in steps.  First, due to the acceleration effect, the detector 
responds with some probability:
\begin{equation}
  \ket{\psi_k(M)} \ket{d_0} 
  \rightarrow \sum_{i = 0}^{i_{\rm max}} \sum_{k'=1}^{n(M)}
    \zeta^k_{k'i}\, \ket{\tilde{\psi}_{k' i}(M)} \ket{d_i},
\label{eq:step-1}
\end{equation}
where we have included the possibility that no response occurs in the 
relevant time interval, represented by $i=0$ in the sum.  The states 
$\ket{\tilde{\psi}_{k' i}(M)}$ arise as a result of the backreaction 
of the detector response, and are not vacuum states.  In the static 
description, the evolution in Eq.~(\ref{eq:step-1}) occurs because the 
detector may absorb particles from the thermal bath; in the inertial 
description, it represents the (probabilistic) emission of particles 
from the detector, accompanied by the internal excitation.

It is important that the coefficients $\zeta^k_{k'i}$ in 
Eq.~(\ref{eq:step-1}) represents a unitary map from the $k$ space 
to its image in the $(k'i)$ space, not in the $k'$ space.  More 
specifically, for a fixed $i$ the rank of the matrix $(\zeta^i)_{kk'} 
\equiv \zeta^k_{k'i}$ in the $k$-$k'$ space is $n(M_i)$, which is 
smaller than $n(M)$, where $M_i$ is given in Eq.~(\ref{eq:probe}). 
This is because in the static description, which we adopt here, the 
excitations in the states $\ket{\tilde{\psi}_{k'i}(M)}$ have negative 
energy resulting from the elimination of particles from the thermal bath, 
so that they carry negative entropy.  In other words, we are assuming 
that only a subspace of the entire space spanned by all the values of 
$k' = 1,\cdots,n(M)$ is realized by a physical process when the state 
is accompanied by negative energy excitations $i$ (which implies 
that the physical Hilbert space for the states having negative energy 
excitations is smaller than that of the naive Fock space suggested by 
the construction in Section~\ref{subsec:interior}).  This dynamical 
assumption allows us to avoid the argument for firewalls presented 
in Ref.~\cite{Almheiri:2013hfa}.  (In the inertial, or infalling, 
description, there is no acceleration induced thermal bath, so that 
the corresponding part of the $k$ index is not accessible locally by 
the detector.  This is consistent because the description in spacetime 
in this case is available only for the timescale of the fall.%
\footnote{We may still suspect that the infalling picture provides a 
 unitarily equivalent description for the distant one because of the 
 existence of the ``horizon'' surrounding the origin, $p_0$, of the 
 infalling reference frame~\cite{Nomura:2013nya} (not to be confused 
 with the horizon as viewed from the distant reference frame). 
 As $p_0$ approaches the singularity, the ``horizon'' approaches 
 $p_0$, eventually making the whole quantum state a ``singularity 
 state''~\cite{Nomura:2011rb}; see Section~\ref{sec:discuss}.  This 
 allows us to contemplate the possibility that the fate of the information 
 mined by the detector is mapped to the intrinsically quantum gravitational 
 dynamics of singularity states after $p_0$ hits the singularity.})

The excitations in $\ket{\tilde{\psi}_{k'i}(M)}$ are ``emitted'' from 
the detector.  These excitations propagate within the causal future of 
the detector response event, and most of them will interact with the 
black hole degrees of freedom, in particular the stretched horizon. 
This results in the evolution of the states
\begin{equation}
  \ket{\tilde{\psi}_{k'i}(M)} \rightarrow 
    \sum_{k_i = 1}^{n(M_i)} \eta^{k'i}_{k_i}\, \ket{\psi_{k_i}(M_i)},
\label{eq:step-2}
\end{equation}
where we have made the simplifying assumption that all the excitations 
interact with the black hole degrees of freedom, and the resulting 
states are fully relaxed into (superpositions of) vacuum states. 
Substituting Eq.~(\ref{eq:step-2}) into Eq.~(\ref{eq:step-1}), we 
obtain the expression in Eq.~(\ref{eq:probe}):
\begin{equation}
  \ket{\psi_k(M)} \ket{d_0} \rightarrow \sum_i \sum_{k_i = 1}^{n(M_i)} 
    \alpha^k_{k_i i} \ket{\psi_{k_i}(M_i)} \ket{d_i},
\label{eq:step-3}
\end{equation}
where $\alpha^k_{k_i i} = \sum_{k'=1}^{n(M)} \zeta^k_{k'i}\, 
\eta^{k'i}_{k_i}$.  Since the distribution of the information about 
the vacuum state is expected to follow the thermal entropy calculated 
in semiclassical field theory, as discussed in Eq.~(\ref{eq:separate'}) 
and below, the rate of information recovery through a mining at $r = 
r_{\rm d}$ will be determined by the thermal entropy at $r = r_{\rm d}$, 
associated with the blueshifted local Hawking temperature.  This implies 
that the mining process, in fact, is expected to achieve the required 
acceleration of the information transfer.

In the semiclassical approximation, the indices such as $k$ and $k'$ 
are not visible because the degrees of freedom represented by these 
indices cannot be resolved.  In practice, this implies that semiclassical 
field theory describes a process as if these indices were absent.  For 
example, it describes the process in Eq.~(\ref{eq:step-1}) as
\begin{equation}
  \ket{\psi(M)} \ket{d_0} 
  \rightarrow \sum_{i = 0}^{i_{\rm max}} 
    \zeta_i\, \ket{\tilde{\psi}_i(M)} \ket{d_i},
\label{eq:step-1-sc}
\end{equation}
where $\ket{\psi(M)}$ is the field theory vacuum state, and 
$\ket{\tilde{\psi}_i(M)}$ represents the state having a negative 
energy excitation labeled by $i$ on $\ket{\psi(M)}$.  This corresponds 
to the description given in Ref.~\cite{Unruh:1983ms}.  The state 
$\ket{\tilde{\psi}_i(M)}$ may be considered to relax into the vacuum 
state $\ket{\psi(M_i)}$ after the excitation interacts with the stretched 
horizon, although this process cannot be described in semiclassical 
field theory.  In any event, in the semiclassical approximation, the 
detector does not have any information about the original black hole 
microstate---the reduced density matrix representing the detector 
state after the response is simply $(\rho)_{ii'} = |\zeta_i|^2 
\delta_{ii'}$, which does not depend on $k$.

It is instructive to consider what happens in the limit $M \rightarrow 
\infty$, in which the relevant spacetime in the static description 
reduces to a Rindler wedge, a portion of Minkowski space.  In this 
case, the entropy density of Eq.~(\ref{eq:S_BH}) diverges, $S = 4\pi 
M^2 l_{\rm P}^2 \rightarrow \infty$, so that we have an infinitely 
large degeneracy of microscopic vacuum states  in any finite energy 
interval.%
\footnote{The exact Minkowski vacuum space may be an artifact of 
 mathematical idealization, and it is possible that in any physically 
 relevant case, there is a boundary/horizon at a finite affine distance 
 in any direction from the origin of the reference frame.  Our argument 
 below persists even in this case by replacing the infinities by 
 appropriately large numbers.}
This implies that to describe any experiment performed in finite time, 
we must coarse-grain an infinite number of microstates labeled by 
the index $k$.  This makes Eq.~(\ref{eq:step-1-sc}) ``exact'' in the 
sense that any finite resolution (or finite time interval) forces us 
to describe the process as in Eq.~(\ref{eq:step-1-sc}), in which the 
vacuum index $k$ does not appear.  In this way, the uniqueness of the 
Minkowski vacuum is recovered, which is crucial for our ability to 
describe physics without having (an infinite amount of) information 
about the specific vacuum we live in.  (Another way to see the uniqueness 
of the Minkowski vacuum, in an inertial description, will be discussed 
in Section~\ref{sec:discuss}.)

\subsection{Effect from spacetime curvature}
\label{subsec:curvature}

We now consider the other effect responsible for the detector response:\ 
the curvature induced effect.  As discussed before, this effect is 
solely responsible for the spontaneous Hawking emission process, where 
the detection of emitted quanta is envisioned only in the asymptotic 
region.  The situation in this case is different from that described above, 
in that the effect does not require an operation of a non-gravitational 
force.  There is therefore no reason to expect that there must be any 
backreaction effect visible in the semiclassical approximation.  In 
fact, we assert that the phenomenon of particle creation due to spacetime 
curvature must be viewed purely as an ``activity of a vacuum,'' whose 
details are invisible in semiclassical field theory.

Consider a process in which Hawking quanta are emitted from the black 
hole to the far exterior region.  (We can imagine measuring these quanta 
by a physical detector at a faraway place, but we are not interested in 
that process here.)  At the microscopic level, the elementary emission 
process occurring in a time interval of $t' - t \approx O(M l_{\rm P}^2)$ 
is given as in Eq.~(\ref{eq:emission-gen}), which, ignoring the difference 
between various $E_i$'s, reads
\begin{equation}
  \ket{\psi_k(M(t))} \ket{\phi_a} \rightarrow 
    \sum_i \sum_{k' = 1}^{n(M(t'))} 
    \beta^k_{k' i} \ket{\psi_{k'}(M(t'))} \ket{\phi_{a+i}}.
\label{eq:el-emit}
\end{equation}
We assert that the states $\ket{\psi_{k'}(M(t'))}$ appearing here 
are vacuum states from the viewpoint of semiclassical field theory. 
In particular, when we act a field theory operator of the form in 
Eqs.~(\ref{eq:modes},~\ref{eq:modes-2}) which has a support in a near 
horizon region, then all these states respond as the thermofield double 
state of the form in Eq.~(\ref{eq:psi_k}).

Let us now discuss a physical picture behind the information extraction 
processes in Eq.~(\ref{eq:el-emit}) and in Eq.~(\ref{eq:step-1}). 
The black hole vacuum states $\ket{\psi_k(M)}$ represent the set of 
microscopic quantum states which, in the semiclassical approximation, 
all look like the vacuum state in the background of a black hole 
of mass $M$ in a fixed location.  While invisible in semiclassical 
gravity, these $\ket{\psi_k(M)}$'s have nontrivial structures 
representing subtle quantum fluctuations of the spacetime beyond 
the resolution of the semiclassical approximation.  The expressions 
in Eqs.~(\ref{eq:step-1},~\ref{eq:el-emit}) imply that the backreactions 
on the black hole relevant for the information transfer go into these 
subtle quantum modes associated with the spacetime.  (Note that the 
excitation $i$ of $\ket{\tilde{\psi}_{k' i}(M)}$ in Eq.~(\ref{eq:step-1}) 
is not directly relevant for the information transfer as can be seen 
in Eq.~(\ref{eq:step-1-sc}).)  These backreactions are not visible 
in semiclassical field theory---they are ``too soft'' to be resolved. 
In particular, the emission of a Hawking quantum must be viewed at 
the semiclassical level as occurring around the barrier region of 
the effective gravitational potential, with the associated information 
transfer occurring through the delocalization of the black hole 
information to the entire zone.

The discussion above says that one must be careful in applying intuition 
from semiclassical field theory to the black hole information release 
processes.  In Hawking emission, the purifiers of the emitted Hawking 
quanta are what semiclassical gravity describes as vacuum states.  In 
particular, they are not field theory modes associated with operators 
$b_{\sigma^+}^{(k)}$ and $b_{\sigma^+}^{(k)\dagger}$ as envisioned in 
Refs.~\cite{Almheiri:2012rt,Almheiri:2013hfa}.  A similar statement 
also applies to the mining process; the purifiers of the modes mined 
from black hole degrees of freedom are black hole vacuum states.  The 
microscopic details of the information transfer processes, in particular 
the flow of information within the black hole degrees of freedom, are 
not visible in semiclassical field theory; one can only see certain 
inclusive quantities that can be calculated as expectation values of 
field theory operators in the vacuum, e.g.\ the total energy flux over 
a large surface~\cite{Christensen:1977jc}.  If one wants to include 
the backreaction on spacetime at the semiclassical level, the best one 
can do seems to do it ``by hand'' such that the semiclassical Einstein 
equation, $R_{\mu\nu} - (1/2) g_{\mu\nu} R = 8\pi l_{\rm P}^2 \langle 
T_{\mu\nu} \rangle$, is satisfied~\cite{Bardeen:1981zz}, which employs 
the Vaidya metric near the horizon.

\subsection{Horizon of an evaporating black hole}
\label{subsec:horizon}

Consider describing the formation and evaporation of a black hole in 
a distant reference frame.  At the microscopic level, the whole process 
is described as a unitary evolution:%
\footnote{As before, we ignore macroscopic dispersions of the black hole 
 mass and location, which may be included explicitly if one wants.}
\begin{equation}
  \ket{m_{\rm init}} \,\,\rightarrow\,\, 
    \sum_{k=1}^{n(M(t))}\! c_k(t)\, \ket{\psi_k(M(t))}\, \ket{r_k(t)} 
  \,\,\rightarrow\,\, \ket{r_{\rm fin}},
\label{eq:BH-evol}
\end{equation}
where $\ket{m_{\rm init}}$, $\ket{r_k(t)}$, and $\ket{r_{\rm fin}}$ 
represent the states for the initial collapsing matter, the subsystem 
complement to the black hole at time $t$ (which includes Hawking radiation 
emitted earlier), and the final Hawking quanta after the black hole 
is completely evaporated.  For generic initial states and microscopic 
emission dynamics, this evolution indeed satisfies the behavior outlined 
by Page~\cite{Page:1993wv} on general grounds.

We here highlight two important aspects of the evolution that have not 
been stated explicitly:
\begin{itemize}
\item[(i)]
{\it Microscopic dynamics need not violate locality at large distances 
in the description based on a distant reference frame} 
--- Since later Hawking emissions occur within the causal future of the 
region in which an earlier emission took place, there is no reason to 
expect that the information recovery process must violate locality, or 
causality, of the spacetime structure.  Namely, while the detailed flow 
of information is not visible to semiclassical field theory, the speed 
at which physical information is transferred can still be bounded by 
the causal structure of the spacetime.
\item[(ii)]
{\it The semiclassical description of the evolution violates unitarity 
at all stages of the black hole formation and evaporation processes} 
--- First, since the same semiclassical black hole (within the 
uncertainties needed to localize it in spacetime) can be formed by 
collapsing matter in different initial states, the description of the 
black hole formation is not unitary in semiclassical gravity.  After 
the formation, the microscopic dynamics makes the black hole emit Hawking 
quanta through the unitary process in Eq.~(\ref{eq:emission-gen}) that 
occurs successively in each time interval of order $M(t) l_{\rm P}^2$. 
Semiclassical gravity, however, describes this as a succession of 
the process in Eq.~(\ref{eq:emission-mixed-2}); in particular, the 
state of the emitted Hawking quanta in each time interval is given 
by the incoherent thermal superposition, making the final Hawking 
radiation state a mixed thermal state.  The semiclassical description 
of a process involving ``microscopic degrees of freedom of spacetime'' 
(the index $k$) is fundamentally non-unitary.
\end{itemize}

We now ask what a physical object will find if it falls into the horizon 
of an evaporating black hole.  At the microscopic level, the state of 
a black hole is in general given by
\begin{equation}
  \rho_{\rm BH}(M(t)) 
  = \sum_{k,l=1}^{n(M(t))}\! c_k(t) c_l^*(t) \, 
    \ket{\psi_k(M(t))}\, \bra{\psi_l(M(t))},
\label{eq:BH-mixed}
\end{equation}
obtained by integrating out the rest of the system in the middle expression 
in Eq.~(\ref{eq:BH-evol}).  As discussed in Section~\ref{subsec:curvature}, 
the states $\ket{\psi_k(M(t))}$ are all vacuum states in the sense that 
their responses to the field theory operators are those of the vacuum 
state in semiclassical field theory.  If a physical object is falling 
into the black hole, it is represented as excitations on the background:\ 
$a_\xi^\dagger$, $b_{\sigma^-}^\dagger$, and $\tilde{b}_{\sigma^-}^\dagger$ 
acting on the microstate.  What happens to the object afterward?

Since $a_\xi^\dagger$ consists of a linear combination of operators 
in the both exterior regions, $b_{\sigma^+}, b_{\sigma^+}^\dagger$ 
and $\tilde{b}_{\sigma^+}, \tilde{b}_{\sigma^+}^\dagger$, the excitations 
in general involve entanglement between those in the two regions.  One 
might think this violates causality because the object that has just 
entered the zone already involves ``excitations'' in the second exterior 
region generated by $\tilde{b}_{\sigma^+}, \tilde{b}_{\sigma^+}^\dagger$. 
This apparent violation, however, is not physical---exciting a black hole 
vacuum state by mode operators in the first or second region, by itself, 
does not have an invariant meaning (as can be seen, e.g., from the fact 
that operating $b_{\sigma^+}^\dagger$ and $\tilde{b}_{\sigma^+}$ lead 
to the same state), so that the physical excitations are still confined 
in the first exterior region near the edge of the zone.  The entangled 
nature of the excitations, however, becomes important when the object 
enters the horizon.

Because the algebra of the operators representing the object and the 
structure of the (vacuum) states on which they act are the same as in 
the semiclassical field theory in the regime of their validity indicated 
by Eq.~(\ref{eq:S_excit}), the only condition needed for the object 
to smoothly pass the horizon is that the dynamics is also the same. 
Therefore, assuming that the microscopic dynamics describing the fallen 
object is well approximated by the Hamiltonian of the quantum field 
theory form in the relevant timescale for the fall:
\begin{equation}
  H(M(t)) \approx 
    H_{\rm QFT}\bigl( a_\xi, b_{\sigma^-}, \tilde{b}_{\sigma^-}; M(t) \bigr) 
  = \sum_k H_{\rm QFT}\bigl( a_\xi^{(k)}, b_{\sigma^-}^{(k)}, 
      \tilde{b}_{\sigma^-}^{(k)}; M(t) \bigr),
\label{eq:H_QFT-fin}
\end{equation}
we can conclude that the object does not see anything special at the 
horizon; in particular, it does not see a violation of the equivalence 
principle there.  Of course, Eq.~(\ref{eq:H_QFT-fin}) is not directly 
derived from the microscopic theory of quantum gravity, but it 
is a well motivated assumption based on the success of general 
relativity.  In fact, we do not find any inconsistency in postulating 
it; in particular, we do not find a necessity of introducing a 
drastically new physical effect such as the firewall discussed in 
Refs.~\cite{Almheiri:2012rt,Braunstein:2009my}.  If we mine the black 
hole by a physical detector, a part of its backreaction is visible to 
a semiclassical observer.  This effect, however, is confined in the 
causal future of the mining event, and is caused by the (non-gravitational) 
force supporting the detector; this is not the firewall phenomenon.

\section{Discussions}
\label{sec:discuss}

In this paper, we have discussed relations between the microscopic 
structure of quantum gravity and what is called the semiclassical 
approximation.  In particular, we have discussed what semiclassical 
field theory really is, and how its salient features---including 
the existence of the interior of a black hole---might arise from the 
fundamental theory of quantum gravity.

This emergence of the semiclassical description by no means implies 
that semiclassical field theory is always valid in the regime in which 
it is conventionally thought to be valid.  As is well known, the global 
spacetime picture of general relativity contradicts the assumption that 
the final state of Hawking evaporation carries the complete information 
about the initial state~\cite{Preskill:1992tc}, the assumption also 
suggested by gauge/gravity duality~\cite{Maldacena:1997re}.  To prevent 
the resulting inconsistency, the fundamental (presumably covariant) 
theory of quantum gravity must possess certain nonlocality beyond 
cutting off the ultraviolet region of semiclassical field theory. 
To explore further aspects of quantum gravity, our discussions must 
be put in this larger context.

In this section, we expand our discussions in more general contexts 
in quantum gravity.  We first present relatively straightforward 
extensions of our analyses to de~Sitter and Minkowski spaces.  We then 
discuss how our picture applies in understanding complementarity, both 
in black hole and more general spacetimes.  Finally, we discuss how 
all these ingredients might help to understand cosmology, especially 
the issue of the beginning/end of the eternally inflating multiverse.

The discussions in this section become more conjectural as we move 
forward.  This is especially so in the discussions of the last two 
subjects---complementarity and the multiverse---which are mostly based 
on the picture one of the authors (Y.N.) has been promoting in the 
past few years.  Our hope is that they set a ground for, or provide 
useful intuition for, further explorations of quantum gravity, which 
may eventually lead to a ``complete understanding'' of the world 
we live in.

\subsubsection*{Simple extensions --- de~Sitter and Minkowski spaces}

Our analyses can be extended relatively straightforwardly to de~Sitter 
space.  Consider de~Sitter space with the Hubble radius $\alpha$.  In 
cosmology, de~Sitter space appears as a meta-stable state in the middle 
of the evolution of the universe, and in this sense it is similar to 
spacetimes with a black hole.  Indeed, string theory suggests that there 
is no absolutely stable de~Sitter vacuum state in full quantum gravity---it 
must decay, at least, before the recurrence time~\cite{Kachru:2003aw}. 
This implies that what we call de~Sitter space cannot be an eigenstate 
of energy (at least in this context).

Suppose we want to determine the time at which the de~Sitter space 
is created or decays within an uncertainty of order $\delta t \approx 
\alpha$, where $t$ is taken as the proper time for a static observer 
(see later for more discussions on this point). This implies the 
existence of uncertainty in the corresponding energy of order 
$\delta E \approx 1/\alpha$.  Estimating this energy to be of order 
the vacuum energy $\rho_\Lambda$ integrated over a Hubble volume, 
$E \approx O(\rho_\Lambda \alpha^3) \approx O(\alpha/l_{\rm P}^2)$, 
this uncertainty is translated into
\begin{equation}
  \delta \alpha \approx O\biggl(\frac{l_{\rm P}^2}{\alpha}\biggr).
\label{eq:delta-alpha}
\end{equation}
As in the case for black holes, we may consider the Gibbons-Hawking 
entropy~\cite{Gibbons:1977mu}
\begin{equation}
  S_{\rm dS} = \frac{{\cal A}_{\rm dS}}{4 l_{\rm P}^2}
  = \frac{\pi \alpha^2}{l_{\rm P}^2},
\label{eq:S_dS}
\end{equation}
to be the entropy density for de~Sitter vacuum states; namely, the 
number of de~Sitter vacuum states with the Hubble radius between 
$\alpha$ and $\alpha + \delta \alpha$ is given by $\approx e^{S_{\rm dS}} 
\delta \alpha/\alpha$.  Here, ${\cal A}_{\rm dS} = 4\pi \alpha^2$ is the 
area of the de~Sitter horizon, and the expression in Eq.~(\ref{eq:S_dS}) 
is understood to be valid at the leading order in expansion in powers 
of $l_{\rm P}/\alpha$.  It is easy to see that the entropy density of 
Eq.~(\ref{eq:S_dS}), in fact, does not depend on the precise choice of 
$\delta \alpha$ as long as $\delta \alpha/\alpha \gg e^{-S_{\rm dS}}$.

Using static coordinates, appropriate for a static observer, the 
de~Sitter horizon is located at the radius $r = \alpha$, and the 
stretched horizon is at $r_{\rm s} = \alpha - O(l_{\rm P}^2/\alpha)$, 
where we have identified the string (cutoff) scale, $l_{\rm s}$, with 
the Planck scale, $l_{\rm P}$.  When a physical detector is located 
at $r = r_{\rm d}$, it sees the blue-shifted local Gibbons-Hawking 
temperature
\begin{equation}
  T_{\rm dS}(r_{\rm d}) 
  = \frac{1}{2\pi\alpha} \frac{1}{\sqrt{1-(r_{\rm d}/\alpha)^2}}.
\label{eq:T_dS}
\end{equation}
The interpretation of this phenomenon goes as in the black hole case. 
If the detector is held at a constant radius $r_{\rm d}$ which is close 
to the horizon, then most of the detector response arises because of 
the acceleration of the detector with respect to local inertial frames 
there.  (This effect by itself would give a response with temperature 
$T_{\rm U}(r_{\rm d}) = r_{\rm d} T_{\rm dS}(r_{\rm d}) / \alpha$.) 
The energy responsible for the response of the detector as well as 
the backreaction disturbing the vacuum comes from the external force 
needed to hold the detector at a constant $r = r_{\rm d}$ for time 
interval of order $\varDelta t \approx \alpha$, as measured at $r=0$. 
This acceleration induced effect, however, does not fully explain 
Eq.~(\ref{eq:T_dS}); in particular, it cannot make a detector located 
at $r = 0$ click as implied by Eq.~(\ref{eq:T_dS}).  As in the case of 
black holes, the remaining effect comes from spacetime curvature, which 
leads to the uncertainty of the local vacuum energy density of order 
$(1/2\pi\alpha)^4$ throughout the Hubble volume.  It is this curvature 
induced effect that is responsible for the response of inertial detectors, 
including the detector located at $r = 0$.

If we denote the de~Sitter vacuum states with the Hubble radius between 
$\alpha$ and $\alpha + \delta \alpha$ by $\ket{\Psi_k(\alpha)}$, the 
``vacuum index'' $k$ runs over
\begin{equation}
  k = 1, \cdots, e^{S_{\rm dS}} = e^{\frac{\pi \alpha^2}{l_{\rm P}^2}}.
\label{eq:dS-k}
\end{equation}
As in the black hole case, most of this information will be in the region 
close to the horizon $r = \alpha$, although some of the information must 
spread over the entire Hubble volume.  The expected distribution follows 
that of the thermal entropy calculated in semiclassical field theory using 
the local temperature in Eq.~(\ref{eq:T_dS}).  The existence of excitations 
in the interior or near exterior region is expected to provide only a small 
perturbation in terms of the entropy counting.

Given the absence of evidence otherwise, we could expect that the 
stretched horizon of de~Sitter space behaves similarly to that of 
a black hole.  When viewed from a static reference frame, a basis 
vector of the Hilbert space ${\cal H}_{\tilde{\Psi}_k}$, spanned by 
$\ket{\Psi_k(\alpha)}$ and its excited states for fixed $k$, will 
be specified by the state of the stretched horizon, $\tilde{\imath}$, 
as well as the number of excitations $n_\sigma$ in each field theory 
mode $\sigma$ inside the stretched horizon:
\begin{equation}
  \ket{\tilde{\Psi}_{k;\tilde{\imath} \{ n_\sigma \}}(\alpha)} 
  = \ket{\tilde{\imath}; k} \ket{\{ n_\sigma \}; k}.
\label{eq:dS-basis}
\end{equation}
Here, the meaning of the index $k$ must be understood similarly to 
the black hole case in Eq.~(\ref{eq:basis}); see the discussion in 
the paragraph containing Eqs.~(\ref{eq:basis-2}~--~\ref{eq:basis-4}). 
While it is not fully clear how the region outside the horizon is encoded 
in general in a given state, we may expect in analogy with the black hole 
case that some of the stretched horizon degrees of freedom are organized 
such that they can be viewed as the mirror modes for $\sigma$.  The 
vacuum states will then take the form
\begin{equation}
  \ket{\Psi_k(\alpha)} \approx 
    \frac{1}{\sqrt{Z}} \sum_{\{ n_\sigma \}} 
    e^{-\frac{E_{\{ n_\sigma \}}}{2 T_{\rm dS}}} 
    \ket{\{ \tilde{n}_\sigma = n_\sigma \}; k} 
    \ket{\{ n_\sigma \}; k};
\qquad
  Z = \sum_{\{ n_\sigma \}} 
    e^{-\frac{E_{\{ n_\sigma \}}}{T_{\rm dS}}},
\label{eq:dS_Psi_k}
\end{equation}
in the near horizon region.  Here, $T_{\rm dS} = 1/2\pi\alpha$, and 
$E_{\{ n_\sigma \}}$ is the energy of the state $\ket{\{ n_\sigma \}; k}$ 
as defined at $r=0$.  This structure will ensure that the de~Sitter 
horizon is smooth.

When an object hits the horizon in a static description, it can be 
thought of as going to space outside the horizon.  The information 
about the object that goes outside will be stored in the state, consisting 
of the interior and the stretched horizon degrees of freedom (although 
there will be some maximum capacity for this information content). 
We expect that such information can be recovered later; otherwise, 
there is no Poincar\'{e} recurrence.  This information recovery may not 
be necessarily in the form of Hawking radiation if the system evolves, 
for example, into Minkowski space or another de~Sitter space with a 
smaller vacuum energy.  Indeed, this is believed to have happened to 
density fluctuations generated in the early inflationary phase in our 
own universe~\cite{Hawking:1982cz}.

The situation in Minkowski space is obtained by taking the limit 
$\alpha \rightarrow \infty$.  In this limit, the ``horizon'' area 
becomes infinity, ${\cal A}_{\rm Min} \rightarrow \infty$, so that 
the number of vacuum states also becomes infinity, $e^{S_{\rm Min}} 
\rightarrow \infty$.  Since the ``horizon'' is located at an infinite 
distance, however, a local observer cannot probe which vacuum he/she 
is in by directly interacting with the ``horizon.''  Also, while 
the information about the vacuum is to some extent delocalized, 
the effect the observer can probe locally away from the ``horizon'' 
in any finite time is expected to be suppressed exponentially in 
${\cal A}_{\rm Min}/l_{\rm P}^2$ (and suppressed, at least, by powers 
of ${\cal A}_{\rm Min}/l_{\rm P}^2$), and hence negligible.  The 
uniqueness of the Minkowski vacuum is thus recovered for the purpose 
of describing physics in any finite spacetime region.  Note that this 
provides a way of arriving at the same conclusion in an inertial reference 
frame as the one reached at the end of Section~\ref{subsec:acceleration} 
using an accelerated reference frame.

\subsubsection*{Complementarity}

A naive formulation of local quantum field theory on spacetime in general 
relativity allows us to use arbitrary spacelike (or null) hypersurfaces 
as the quantization surfaces, as long as spacetime curvature is 
sufficiently small everywhere on it.  In black hole physics, this 
allows for a hypersurface that passes through both an object fallen 
inside the horizon as well as late Hawking quanta, contradicting the 
idea that final-state Hawking radiation contains the full information 
about the system that collapsed and fell into the black hole (since 
it leads to a cloning of quantum information)~\cite{Preskill:1992tc}. 
We expect that this problem is avoided in the fundamental theory of 
quantum gravity due to some nonlocality at large distances that becomes 
prominent when certain quantization surfaces are chosen.

The idea of black hole complementarity~\cite{Susskind:1993if,Hayden:2007cs} 
is that we may quantize the system such that this nonlocality is 
``minimized,'' and that there are multiple equivalent ways to do 
so. For example, by restricting the quantization surfaces appropriately 
to foliate only the exterior region, we may obtain a unitary description 
of physics in which the long-range nonlocality effect described above 
is absent.  In addition, there are other ``infalling'' ways to quantize 
the system in which portions of the interior spacetime are included 
in the descriptions.  Our discussions in this paper support this idea. 
In particular, we do not find any fundamental inconsistency between 
the existence of infalling descriptions and unitarity of the Hawking 
evaporation process as viewed from a distant reference frame.

While the general and precise formulation of complementarity is 
not yet known, a simple way to implement it seems to describe a 
system as viewed from a freely falling (local Lorentz) reference 
frame~\cite{Nomura:2011dt,Nomura:2011rb,Nomura:2013nya}.  Specifically, 
we may consider a fixed reference point $p_0$ and a null or spacelike 
hypersurface associated with it, which in general is bounded by 
certain ``horizons'' signaling the breakdown of the local spacetime 
picture.  It is important that the {\it procedure} to determine the 
hypersurface here is given independently of background spacetime; for 
example, it can be generated by past-directed light rays emanating 
from $p_0$ in all angular directions, with each light ray terminated 
when a certain condition on its expansion or other quantities is met. 
A quantum state that allows for a semiclassical spacetime interpretation 
is then given by specifying the configuration of physical degrees of 
freedom on the hypersurface using a coordinate system related to a 
local Lorentz frame elected at $p_0$.  In the fundamental theory 
of quantum gravity, restricting quantum states to this kind corresponds 
to partially fixing the gauge associated with spacetime transformations; 
after fixing the coordinates on the hypersurface, the remaining gauge 
freedom corresponds to the choice of the reference frame, whose generators 
we denote by $H$, $P_i$, $J_{[ij]}$, and $K_i$, borrowing the standard 
notation for the Poincar\'{e} algebra.

Let us denote the Hilbert space spanned by all the states obtained as 
described above by ${\cal H}_{\rm spacetime}$.  The relevant, partially 
gauge-fixed Hilbert space ${\cal H}$ is then given by the direct sum of 
${\cal H}_{\rm spacetime}$ and another Hilbert space ${\cal H}_{\rm sing}$ 
containing quantum states that do not allow for a spacetime interpretation 
(which become relevant when $p_0$ hits a spacetime singularity):%
\footnote{Here, we have changed the notation from that in 
 Ref.~\cite{Nomura:2013nya}.  The Hilbert spaces denoted by ${\cal H}$ 
 and ${\cal H}_{\rm spacetime}$ here correspond to ${\cal H}_{\rm QG}$ 
 and ${\cal H}$ in Ref.~\cite{Nomura:2013nya}, respectively.}
\begin{equation}
  {\cal H} = {\cal H}_{\rm spacetime} \oplus {\cal H}_{\rm sing}.
\label{eq:Hilbert_QG}
\end{equation}
Describing a system from a fixed reference frame corresponds to further 
fixing the gauge for the transformations associated with $P_i$, $J_{[ij]}$, 
and $K_i$ by taking an appropriate section of the corresponding gauge 
orbit in ${\cal H}$.  A quantum state representing the system at a 
fixed time is then given by an element in ${\cal H}$, which may involve 
a superposition of macroscopically different spacetimes or even a 
superposition of spacetime and singularity states.  Its evolution is 
given by the (effective) Hamiltonian $H_{\rm QG}$, generating a translation 
of the proper time $\tau$ measured at $p_0$.  (This transformation 
corresponds to the remaining gauge freedom represented by generator 
$H$.  The emergence of the effective time evolution picture from a more 
fundamental description in quantum gravity will be discussed below.)

In this framework, a complementarity transformation is simply choosing 
a different local Lorentz frame at some time $\tau$, which corresponds 
to choosing a different section in the orbit of the gauge transformations 
generated by $P_i$ and $K_i$ (and $J_{[ij]}$, which however yields only 
a trivial rotation of the frame).  This framework was applied in particular 
to black hole physics in Ref.~\cite{Nomura:2012cx}, in which it was shown 
how the interior spacetime may appear effectively by combining various 
descriptions in different reference frames.

We may take a basis in ${\cal H}_{\rm spacetime}$ in which a basis 
element represents a hypersurface that contains $p_0$ and is surrounded 
by the boundary (horizon) that indicates the breakdown of the semiclassical 
spacetime picture.  For example, if $p_0$ is located outside the 
Schwarzschild horizon of a black hole in de~Sitter space, then $p_0$ 
will ``see'' the black hole horizon in some directions and the de~Sitter 
horizon in the others; specifically, if the hypersurface is generated 
by past-directed light rays emanating from $p_0$, then each light ray 
hits either the (stretched) black hole or de~Sitter horizon, depending 
on the angular direction in which it is emitted.  A general criterion 
for determining the location of the boundary is not fully understood, 
but a possible procedure was described in Ref.~\cite{Nomura:2013nya} 
which applies to certain simple cases.  Now, consider a set of quantum 
states that share the ``same boundary'' $\partial {\cal M}$ in some 
appropriate sense (although the precise general definition of this 
statement is also not yet available).  We may then expect, based on 
the conjectured entropy bound in Ref.~\cite{Bousso:1999xy}, that the 
dimension of the Hilbert space ${\cal H}_{\partial {\cal M}}$ spanned 
by all these states is given by
\begin{equation}
  \ln {\rm dim}\,{\cal H}_{\partial {\cal M}} 
    \approx \frac{{\cal A}_{\partial {\cal M}}}{4 l_{\rm P}^2},
\label{eq:H_dM}
\end{equation}
where ${\cal A}_{\partial {\cal M}}$ is the total area of the boundary; 
in the above example, ${\cal A}_{\partial {\cal M}} = {\cal A}_{\rm BH} 
+ {\cal A}_{\rm dS}$ where ${\cal A}_{\rm BH}$ and ${\cal A}_{\rm dS}$ 
are the areas of the black hole and de~Sitter portions of the boundary. 
In our present context, ${\rm dim}\,{\cal H}_{\partial {\cal M}}$ 
represents the number of independent microstates that must be coarse-grained 
to obtain the semiclassical description on a hypersurface bounded by 
$\partial {\cal M}$.

\subsubsection*{The static quantum multiverse}

By describing a system from a fixed reference frame, we are fixing all 
the relevant gauge redundancies except for that associated with $H$:\ 
translation of the proper time $\tau$ at $p_0$.  In the most fundamental 
treatment of quantum gravity, this residual gauge redundancy must also 
be fixed, which we may do by imposing the constraint on an element 
$\ket{\Phi}$ of ${\cal H}$ in Eq.~(\ref{eq:Hilbert_QG}) (equivalent 
to taking an appropriate section in the corresponding gauge orbit):
\begin{equation}
  H \ket{\Phi} = 0.
\label{eq:H_Psi}
\end{equation}
The effective time evolution picture may then arise as correlations 
between physical configurations of subsystems~\cite{DeWitt:1967yk}. 
Specifically, in $\ket{\Phi}$ we can identify a (small) subsystem as 
the ``clock'' degrees of freedom, and reinterpret the entanglement 
of these degrees of freedom---represented e.g.\ by a set of states 
$\ket{i}$---with the rest of the degrees of freedom---represented e.g.\ 
by a set of states $\ket{\Psi_i}$---as the time evolution of a state 
$\ket{\Psi_i}$ with $i$ playing the role of time.

We emphasize that $\ket{\Phi}$ in Eq.~(\ref{eq:H_Psi}) is the quantum 
state for the {\it entire} system, including all the degrees of 
freedom existing in the world.  If we omit any, even small, subsystem 
from our description, then the resulting state need not satisfy 
Eq.~(\ref{eq:H_Psi}) (hence making the time evolution picture possible). 
In Ref.~\cite{Nomura:2012zb}, it was conjectured that the condition 
of Eq.~(\ref{eq:H_Psi}) together with the normalizability of the state 
in ${\cal H}$, $\inner{\Phi}{\Phi} < \infty$, is enough to select the 
{\it state of the multiverse}.%
\footnote{If the state is not uniquely selected, but multiple discrete 
 states $\ket{\Phi_n}$ ($n = 1,\cdots,N$) are allowed, then we may expect 
 that the state of the multiverse is mixed:\ $\sum_{n=1}^N \ket{\Phi_n} 
 \bra{\Phi_n} / N$.  Our argument below still applies in this case 
 with appropriate modifications.}
In this picture, practically all the systems we observe, including 
the whole universe of our own, are tiny portions of the branches of 
$\ket{\Phi}$, and their ``evolutions'' are well described by
\begin{equation}
  \ket{\Psi(\tau)} \rightarrow \ket{\Psi(\tau + \varDelta \tau)} 
  = e^{-i H \varDelta \tau} \ket{\Psi(\tau)},
\end{equation}
using the proper time $\tau$ at $p_0$.  This is the time evolution of 
a state we have been discussing so far---it is a parametric representation 
of correlations between configurations of subsystems, analogous to a 
parametric representation of a curve on a plane, $(x(\tau), y(\tau))$. 
Note that this does not contradict the fact that the state of the entire 
multiverse is static:\ $e^{-i H \varDelta \tau} \ket{\Phi} = \ket{\Phi}$.

If the state of the multiverse is indeed selected as described above, 
then we may, in principle, predict everything we are allowed to predict 
once the rule of extracting it is given (assuming that the explicit form 
of relevant operators, such as $H$, and their algebra is known).  This 
rule must reduce to the standard Born rule in appropriate circumstances, 
but it will involve ingredients beyond that because the state of the 
multiverse $\ket{\Phi}$ does not have any environment to interact with, 
whose existence is implicitly assumed by the standard Born rule (e.g.\ 
in determining the basis for projections).  In fact, the normalizability 
of $\ket{\Phi}$ would mean that the number of components in $\ket{\Phi}$, 
when expanded in a basis in which locality is manifest, is effectively 
finite, and the rule would have to be formulated in a way that it works 
internally in the finite-dimensional Hilbert space spanned by these 
components.  Since we cannot use an external environment to select 
a measurement basis, such a rule will have to involve the structure 
of operators, rather than just a state.  In our view, this may very 
well comprise one of the most important challenges to obtain a complete 
theory of quantum gravity, about which the current formulation of 
string theory has little to say.

\section*{Acknowledgments}

We thank Raphael Bousso, Ben Freivogel, Daniel Harlow, Juan Maldacena, 
Joseph Polchinski, Jaime Varela, Herman Verlinde, and I-Sheng Yang for 
various conversations during our exploration of this subject.  This 
work was supported in part by the Director, Office of Science, Office 
of High Energy and Nuclear Physics, of the U.S.\ Department of Energy 
under Contract DE-AC02-05CH11231, and in part by the National Science 
Foundation under grant PHY-1214644.


\begin{thebibliography}{99}

\bibitem{Bekenstein:1973ur}
J.~D.~Bekenstein,
Phys.\ Rev.\  D {\bf 7}, 2333 (1973).

\bibitem{Hawking:1974sw}
S.~W.~Hawking,
Commun.\ Math.\ Phys.\  {\bf 43}, 199 (1975)
[Erratum-ibid.\  {\bf 46}, 206 (1976)].

\bibitem{'tHooft:1990fr}
G.~'t Hooft,
Nucl.\ Phys.\ B {\bf 335}, 138 (1990);
C.~R.~Stephens, G.~'t Hooft and B.~F.~Whiting,
Class.\ Quant.\ Grav.\  {\bf 11}, 621 (1994)
[arXiv:gr-qc/9310006].

\bibitem{Susskind:1993if}
L.~Susskind, L.~Thorlacius and J.~Uglum,
Phys.\ Rev.\  D {\bf 48}, 3743 (1993)
[arXiv:hep-th/9306069];
L.~Susskind and L.~Thorlacius,
Phys.\ Rev.\ D {\bf 49}, 966 (1994)
[hep-th/9308100].

\bibitem{Hayden:2007cs}
P.~Hayden and J.~Preskill,
JHEP {\bf 09}, 120 (2007)
[arXiv:0708.4025 [hep-th]].

\bibitem{Almheiri:2012rt}
A.~Almheiri, D.~Marolf, J.~Polchinski and J.~Sully,
JHEP {\bf 02}, 062 (2013)
[arXiv:1207.3123 [hep-th]].

\bibitem{Hawking:1976ra}
S.~W.~Hawking,
Phys.\ Rev.\ D {\bf 14}, 2460 (1976).

\bibitem{Unruh:1976db}
W.~G.~Unruh,
Phys.\ Rev.\ D {\bf 14}, 870 (1976).

\bibitem{Almheiri:2013hfa}
A.~Almheiri, D.~Marolf, J.~Polchinski, D.~Stanford and J.~Sully,
JHEP {\bf 09}, 018 (2013)
[arXiv:1304.6483 [hep-th]];
D.~Marolf and J.~Polchinski,
Phys.\ Rev.\ Lett.\ {\bf 111}, 171301 (2013)
arXiv:1307.4706 [hep-th].

\bibitem{Nomura:2012ex}
Y.~Nomura and J.~Varela,
JHEP {\bf 07}, 124 (2013)
[arXiv:1211.7033 [hep-th]];
Y.~Nomura, J.~Varela and S.~J.~Weinberg,
Phys.\ Rev.\ D {\bf 88}, 084052 (2013)
[arXiv:1308.4121 [hep-th]].

\bibitem{Nomura:2013lia}
Y.~Nomura and S.~J.~Weinberg,
arXiv:1310.7564 [hep-th].

\bibitem{Giddings:2012bm}
S.~B.~Giddings,
Phys.\ Rev.\ D {\bf 85}, 124063 (2012)
[arXiv:1201.1037 [hep-th]];
Phys.\ Rev.\ D {\bf 88}, 064023 (2013)
[arXiv:1211.7070 [hep-th]].

\bibitem{Papadodimas:2012aq}
K.~Papadodimas and S.~Raju,
JHEP {\bf 10}, 212 (2013)
[arXiv:1211.6767 [hep-th]];
Phys.\ Rev.\ Lett.\  {\bf 112}, 051301 (2014)
[arXiv:1310.6334 [hep-th]];
Phys.\ Rev.\ D {\bf 89}, 086010 (2014)
[arXiv:1310.6335 [hep-th]].

\bibitem{Verlinde:2012cy}
E.~Verlinde and H.~Verlinde,
JHEP {\bf 10}, 107 (2013)
[arXiv:1211.6913 [hep-th]];
arXiv:1306.0515 [hep-th];
arXiv:1311.1137 [hep-th].

\bibitem{Maldacena:2013xja}
J.~Maldacena and L.~Susskind,
Fortsch.\ Phys.\  {\bf 61}, 781 (2013)
[arXiv:1306.0533 [hep-th]].

\bibitem{Bousso:2012as}
R.~Bousso,
Phys.\ Rev.\ D {\bf 87}, 124023 (2013)
[arXiv:1207.5192 [hep-th]];
Y.~Nomura, J.~Varela and S.~J.~Weinberg,
JHEP {\bf 03}, 059 (2013)
[arXiv:1207.6626 [hep-th]];
S.~D.~Mathur and D.~Turton,
JHEP {\bf 01}, 034 (2014)
[arXiv:1208.2005 [hep-th]];
Nucl.\ Phys.\ B {\bf 884}, 566 (2014)
[arXiv:1306.5488 [hep-th]];
B.~D.~Chowdhury and A.~Puhm,
Phys.\ Rev.\ D {\bf 88}, 063509 (2013)
[arXiv:1208.2026 [hep-th]];
T.~Banks and W.~Fischler,
arXiv:1208.4757 [hep-th];
arXiv:1305.3923 [hep-th];
R.~Brustein,
Fortsch.\ Phys.\  {\bf 62}, 255 (2014)
[arXiv:1209.2686 [hep-th]];
L.~Susskind,
arXiv:1210.2098 [hep-th];
arXiv:1311.7379 [hep-th];
arXiv:1402.5674 [hep-th];
S.~G.~Avery, B.~D.~Chowdhury and A.~Puhm,
JHEP {\bf 09}, 012 (2013)
[arXiv:1210.6996 [hep-th]];
D.~Harlow and P.~Hayden,
JHEP {\bf 06}, 085 (2013)
[arXiv:1301.4504 [hep-th]];
R.~Brustein and A.~J.~M.~Medved,
JHEP {\bf 09}, 015 (2013)
[arXiv:1305.3139 [hep-th]];
JHEP {\bf 02}, 116 (2014)
[arXiv:1310.5861 [hep-th]];
arXiv:1312.0880 [hep-th];
D.~A.~Lowe and L.~Thorlacius,
Phys.\ Rev.\ D {\bf 88}, 044012 (2013)
[arXiv:1305.7459 [hep-th]];
arXiv:1402.4545 [hep-th];
D.~N.~Page,
arXiv:1306.0562 [hep-th];
M.~Van Raamsdonk,
arXiv:1307.1796 [hep-th];
S.~Lloyd and J.~Preskill,
arXiv:1308.4209 [hep-th];
S.~D.~H.~Hsu,
arXiv:1308.5686 [hep-th];
I.~Ilgin and I.-S.~Yang,
Phys.\ Rev.\ D {\bf 89}, 044007 (2014)
[arXiv:1311.1219 [hep-th]];
B.~Freivogel,
arXiv:1401.5340 [hep-th].

\bibitem{Nomura:2012cx}
Y.~Nomura, J.~Varela and S.~J.~Weinberg,
Phys.\ Rev.\ D {\bf 87}, 084050 (2013)
[arXiv:1210.6348 [hep-th]].

\bibitem{Page:1979tc}
D.~N.~Page,
Phys.\ Rev.\ Lett.\  {\bf 44}, 301 (1980).

\bibitem{'tHooft:1993gx}
G.~'t Hooft,
gr-qc/9310026.

\bibitem{Susskind:1994vu}
L.~Susskind,
J.\ Math.\ Phys.\  {\bf 36}, 6377 (1995)
[arXiv:hep-th/9409089].

\bibitem{Bousso:1999xy}
R.~Bousso,
JHEP {\bf 07}, 004 (1999)
[arXiv:hep-th/9905177];
Rev.\ Mod.\ Phys.\  {\bf 74}, 825 (2002)
[hep-th/0203101].

\bibitem{Nomura}
Y.~Nomura, talk at KITP rapid response workshop 
``Black Holes:\ Complementarity, Fuzz, or Fire?,'' 
KITP Santa Barbara, August 2013.

\bibitem{Page:1976df} 
D.~N.~Page,
Phys.\ Rev.\ D {\bf 13}, 198 (1976).

\bibitem{Israel:1976ur}
W.~Israel,
Phys.\ Lett.\ A {\bf 57}, 107 (1976).

\bibitem{Sorkin}
R.~D.~Sorkin,
in 10th International Conference on General Relativity and Gravitation,
Contributed Papers, vol.~II, 1167,
eds.~B.~Bertotti, F.~de~Felice, and A.~Pascolini 
(Roma, Consiglio Nazionale Delle Ricerche, 1983).

\bibitem{Nomura:2013nya}
Y.~Nomura, J.~Varela and S.~J.~Weinberg,
Phys.\ Lett.\ B {\bf 733}, 126 (2014)
[arXiv:1304.0448 [hep-th]].

\bibitem{Susskind:1994sm}
L.~Susskind and J.~Uglum,
Phys.\ Rev.\ D {\bf 50}, 2700 (1994)
[hep-th/9401070].

\bibitem{Cooperman:2013iqr}
J.~H.~Cooperman and M.~A.~Luty,
arXiv:1302.1878 [hep-th].

\bibitem{Unruh:1983ms}
W.~G.~Unruh and R.~M.~Wald,
Phys.\ Rev.\ D {\bf 29}, 1047 (1984).

\bibitem{Unruh:1982ic}
W.~G.~Unruh and R.~M.~Wald,
Phys.\ Rev.\ D {\bf 25}, 942 (1982).

\bibitem{Brown:2012un}
A.~R.~Brown,
Phys.\ Rev.\ Lett.\  {\bf 111}, 211301 (2013)
[arXiv:1207.3342 [gr-qc]].

\bibitem{Nomura:2011rb}
Y.~Nomura,
Found.\ Phys.\  {\bf 43}, 978 (2013)
[arXiv:1110.4630 [hep-th]].

\bibitem{Christensen:1977jc}
S.~M.~Christensen and S.~A.~Fulling,
Phys.\ Rev.\ D {\bf 15}, 2088 (1977);
P.~Candelas,
Phys.\ Rev.\ D {\bf 21}, 2185 (1980).

\bibitem{Bardeen:1981zz}
J.~M.~Bardeen,
Phys.\ Rev.\ Lett.\  {\bf 46}, 382 (1981);
R.~Balbinot,
Class.\ Quant.\ Grav.\  {\bf 1}, 573 (1984).

\bibitem{Page:1993wv}
D.~N.~Page,
Phys.\ Rev.\ Lett.\  {\bf 71}, 3743 (1993)
[hep-th/9306083].

\bibitem{Braunstein:2009my}
See also,
S.~L.~Braunstein,
arXiv:0907.1190v1 [quant-ph];
S.~D.~Mathur,
Class.\ Quant.\ Grav.\  {\bf 26}, 224001 (2009)
[arXiv:0909.1038 [hep-th]];
S.~B.~Giddings,
Class.\ Quant.\ Grav.\  {\bf 28}, 025002 (2011)
[arXiv:0911.3395 [hep-th]].

\bibitem{Preskill:1992tc}
For a review,
J.~Preskill,
in {\it Blackholes, Membranes, Wormholes and Superstrings},
ed.\ S.~Kalara and D.~V.~Nanopoulos (World Scientific, Singapore, 1993) p.~22
[hep-th/9209058].

\bibitem{Maldacena:1997re}
J.~Maldacena,
Adv.\ Theor.\ Math.\ Phys.\  {\bf 2}, 231 (1998)
[hep-th/9711200].

\bibitem{Kachru:2003aw}
S.~Kachru, R.~Kallosh, A.~Linde and S.~P.~Trivedi,
Phys.\ Rev.\ D {\bf 68}, 046005 (2003)
[hep-th/0301240].

\bibitem{Gibbons:1977mu}
G.~W.~Gibbons and S.~W.~Hawking,
Phys.\ Rev.\ D {\bf 15}, 2738 (1977).

\bibitem{Hawking:1982cz}
S.~W.~Hawking,
Phys.\ Lett.\ B {\bf 115}, 295 (1982);
A.~A.~Starobinsky,
Phys.\ Lett.\ B {\bf 117}, 175 (1982);
A.~H.~Guth and S.-Y.~Pi,
Phys.\ Rev.\ Lett.\  {\bf 49}, 1110 (1982);
see also,
V.~F.~Mukhanov and G.~V.~Chibisov,
JETP Lett.\  {\bf 33}, 532 (1981)
[Pisma Zh.\ Eksp.\ Teor.\ Fiz.\  {\bf 33}, 549 (1981)].

\bibitem{Nomura:2011dt}
Y.~Nomura,
JHEP {\bf 11}, 063 (2011)
[arXiv:1104.2324 [hep-th]].

\bibitem{DeWitt:1967yk}
B.~S.~DeWitt,
Phys.\ Rev.\  {\bf 160}, 1113 (1967).

\bibitem{Nomura:2012zb}
Y.~Nomura,
Phys.\ Rev.\ D {\bf 86}, 083505 (2012)
[arXiv:1205.5550 [hep-th]].

\end{thebibliography}
\end{document}